\documentclass[draftclsnofoot, 12pt, onecolumn]{IEEEtran}

\usepackage{graphicx,epsf,psfrag}
\usepackage{amsmath,amssymb}
\usepackage{amsfonts}
\usepackage{mathrsfs}
\usepackage{etoolbox}
\usepackage{cite}
\usepackage{bbm}
\usepackage{latexsym}

\newcommand{\beq}{\begin{equation}}
\newcommand{\eeq}{\end{equation}}
\newcommand{\be}{\begin{equation}}
\newcommand{\ee}{\end{equation}}
\newcommand{\eps}{\epsilon}

\newcommand{\bi}{\begin{itemize}}
\newcommand{\ei}{\end{itemize}}
\newcommand{\m}{\item}

\newcommand{\calG}{\mathcal{G}}

\newcommand{\calN}{\mathcal{N}}

\newcommand{\calT}{\mathcal{T}}

\newcommand{\calV}{\mathcal{V}}

\newcommand{\bg}{\mathbf{g}}
\newcommand{\bG}{\mathbf{G}}

\newcommand{\bs}{\mathbf{s}}
\newcommand{\bS}{\mathbf{S}}

\newcommand{\bV}{\mathbf{V}}

\newcommand{\bx}{\mathbf{x}}
\newcommand{\bX}{\mathbf{X}}
\newcommand{\by}{\mathbf{y}}
\newcommand{\bY}{\mathbf{Y}}

\newcommand{\bbE}{\mathbb{E}}

\newcommand{\bbP}{\mathbb{P}}

\newcommand{\bbR}{\mathbb{R}}

\newcommand{\scU}{\mathscr{U}}

\DeclareMathAlphabet{\mathbsf}{OT1}{cmss}{bx}{n}
\DeclareMathAlphabet{\mathssf}{OT1}{cmss}{m}{sl}

\DeclareSymbolFont{bsfletters}{OT1}{cmss}{bx}{n}  
\DeclareSymbolFont{ssfletters}{OT1}{cmss}{m}{n}
\DeclareMathSymbol{\bsfGamma}{0}{bsfletters}{'000}
\DeclareMathSymbol{\ssfGamma}{0}{ssfletters}{'000}
\DeclareMathSymbol{\bsfDelta}{0}{bsfletters}{'001}
\DeclareMathSymbol{\ssfDelta}{0}{ssfletters}{'001}
\DeclareMathSymbol{\bsfTheta}{0}{bsfletters}{'002}
\DeclareMathSymbol{\ssfTheta}{0}{ssfletters}{'002}
\DeclareMathSymbol{\bsfLambda}{0}{bsfletters}{'003}
\DeclareMathSymbol{\ssfLambda}{0}{ssfletters}{'003}
\DeclareMathSymbol{\bsfXi}{0}{bsfletters}{'004}
\DeclareMathSymbol{\ssfXi}{0}{ssfletters}{'004}
\DeclareMathSymbol{\bsfPi}{0}{bsfletters}{'005}
\DeclareMathSymbol{\ssfPi}{0}{ssfletters}{'005}
\DeclareMathSymbol{\bsfSigma}{0}{bsfletters}{'006}
\DeclareMathSymbol{\ssfSigma}{0}{ssfletters}{'006}
\DeclareMathSymbol{\bsfUpsilon}{0}{bsfletters}{'007}
\DeclareMathSymbol{\ssfUpsilon}{0}{ssfletters}{'007}
\DeclareMathSymbol{\bsfPhi}{0}{bsfletters}{'010}
\DeclareMathSymbol{\ssfPhi}{0}{ssfletters}{'010}
\DeclareMathSymbol{\bsfPsi}{0}{bsfletters}{'011}
\DeclareMathSymbol{\ssfPsi}{0}{ssfletters}{'011}
\DeclareMathSymbol{\bsfOmega}{0}{bsfletters}{'012}
\DeclareMathSymbol{\ssfOmega}{0}{ssfletters}{'012}

\DeclareMathOperator*{\argmin}{arg\,min}

\DeclareMathOperator{\var}{Var}

\DeclareMathOperator{\cov}{Cov}

\ifcsmacro{theorem}{}{
\newtheorem{theorem}{Theorem}
\newtheorem{lemma}[theorem]{Lemma}

}

\newcommand{\qednew}{\nobreak \ifvmode \relax \else
      \ifdim\lastskip<1.5em \hskip-\lastskip
      \hskip1.5em plus0em minus0.5em \fi \nobreak
      \vrule height0.75em width0.5em depth0.25em\fi}

\usepackage{enumitem}
\usepackage{bbm}
\usepackage{cite}
\usepackage[usenames, dvipsnames]{color}
\usepackage{soul}
\usepackage{hyperref}
 
\usepackage{subfig}
\usepackage{graphicx}

\allowdisplaybreaks

\IEEEoverridecommandlockouts

\title{\LARGE \bf
Capacity of Gaussian \\
Arbitrarily-Varying Fading Channels}

\author{Fatemeh Hosseinigoki$^*$ and Oliver Kosut$^*$
\thanks{This material is based upon work supported by the National Science Foundation under Grant No. CCF-1453718.}
\thanks{This paper was presented in part at the Conference on Information Sciences and Systems (CISS 2019) \cite{CISSFHOK}.}
\thanks{$^*$School of Electrical, Computer and Energy Engineering, Arizona State University, Tempe, AZ 85287.
        {\tt\small \{fhossei1,okosut\}@asu.edu}}}

\begin{document}
\maketitle

\newcommand{\T}{\calT_{\eps}^{(n)}}

\begin{abstract}
	This paper considers an arbitrarily-varying fading channel consisting of one transmitter, one receiver and an arbitrarily varying adversary. The channel is assumed to have additive Gaussian noise and fast fading of the gain from the legitimate user to the receiver. We study four variants of the problem depending on whether the transmitter and/or adversary have access to the fading gains; we assume the receiver always knows the fading gains. In two variants the adversary does not have access to the gains, we show that the capacity corresponds to the capacity of a standard point-to-point fading channel with increased noise variance. The capacity of the other two cases, in which the adversary has knowledge of the channel gains, are determined by the worst-case noise variance as a function of the channel gain subject to the jammer's power constraint; if the jammer has enough power, then it can imitate the legitimate user's channel, causing the capacity to drop to zero. We also show that having the channel gains causally or non-causally at the encoder and/or the adversary does not change the capacity, except for the case where all parties know the channel gains. In this case, if the transmitter knows the gains non-causally, while the adversary knows the gains causally, then it is possible for the legitimate users to keep a secret from the adversary. We show that in this case the capacity is always positive.\\

\end{abstract}

\makeatletter
\newcommand{\BBigg}{\bBigg@{3.0}}
\newcommand{\vast}{\bBigg@{3.5}}
\newcommand{\vastt}{\bBigg@{4.0}}
\newcommand{\Vast}{\bBigg@{4.5}}
\makeatother

\section{Introduction}\label{1}

It is a matter of great importance to study wireless communication channels since they include numerous challenges caused by noise and fading. The wireless environment also allows any uninvited signal to enter the channel. These signals can act as interference or in the worse case as a malicious jammer whose aim is to disrupt the communication between the legitimate users. In this work, we explore how these various signals interact with each other to restrict the overall capacity of the channel. 

Goldsmith and Varaiya in \cite{Varaiya} studied the point-to-point Gaussian channel with fast fading in which the fading coefficients are drawn i.i.d. from a distribution known to all parties. They determined the capacity of the channel where the fading coefficients are available at the receiver and possibly the transmitter. When the fading coefficients are not available at the transmitter, the capacity is equal to the expected value of the capacity of the corresponding Gaussian channel with the received signal-to-noise ratio affected by the fading gains. If both the transmitter and receiver know the exact channel gains, the transmitter can maximize the capacity by constructing its signal power as a function of the channel gains.

Another line of work studies channels in the presence of a malicious adversary. The adversary can be an active attacker who sends its signal to the channel in order to cease or restrict the communication between the legitimate users. If the adversary's signal is arbitrarily chosen or is drawn from an unknown distribution, then the channel is called as an arbitrary-varying channel (AVC). We focus on the variant of the problem wherein the adversary's knowledge is limited to the code of the legitimate user, but it does not have access to the user's transmitted messages, neither exact nor noisy. Csisz\'ar and Narayan established the capacity of discrete AVC with input and jammer power constraints in \cite{AVCNarayan,AVCNarayan2}. They also derived the capacity for a continuous version of the AVC with input and state (jammer) power constraints in \cite{Csiszar}.

Csisz\'ar and Narayan in \cite{Csiszar} characterized the capacity of the Gaussian arbitrarily-varying channel under the average probability of error and the average power constraints for input and state. It is assumed that the adversary does not have any information about the legitimate signal except the code. It is shown that if the adversary has enough power to forge a message and send it to the channel, then the receiver gets confused and cannot distinguish between the true message and the malicious one. This occurrence is called \emph{symmetrizability} , causing the capacity to drop to zero. In  \cite{Csiszar}, it is shown that the adversary can symmetrize the channel if and only if it has greater power that the legitimate transmitter. However, if the jammer does not have enough power, then the capacity is equal to the capacity of a standard Gaussian channel with the noise variance increased by the power of the jammer.

The problem of AVC capacity has been also studied for discrete and continuous multi-user channels. The authors in \cite{144713} considered the discrete arbitrarily-varying multiple-access memoryless channel. We characterized lower and upper bounds for the capacity of arbitrarily-varying Gaussian interference channel in \cite{FatemehAllerton2016}. List-decoding is also investigated for the discrete and Gaussian AVCs in \cite{Hughes,6157083} and in \cite{FatemehISIT2018}, respectively. The list capacity is derived by using list-decoding which decodes a list of messages instead of unique message at the receiver.

The three elements of Gaussian noise, fading, and adversary have been previously combined to study problems with a (passive) eavesdropping adversary, rather than an (active) transmitting adversary. A secrecy capacity problem with slow fading, in which the fading gains are constant over each block length, is considered in \cite{EavsFading}. The authors determined the secrecy capacity of the channel where both of transmitter and receiver know the channel state information (CSI) of the main path, but do not have any information about the eavesdropper channel. Furthermore, the capacity is generalized to multiple eavesdroppers in \cite{MultiEavsFading}. The problem is also studied in \cite{RayleighFading} for a specific case of a fast Rayleigh fading eavesdropper channel and a standard Gaussian main channel where the CSI are only known to the eavesdropper.

In this paper, we consider a Gaussian AVC with fast fading on the main path, as illustrated in Fig. \ref{fig:GAVFC}; we refer to this channel as the Gaussian arbitrarily-varying fading channel (GAVFC). We characterize the capacity of the GAVFC under the average probability of error criterion. Similar to the Gaussian fading channel, we also assume that everyone knows the fading gain distribution including the adversary, but they may or may not know the realization of the gain sequence. Note that the ``arbitrarily-varying'' aspect of the channel is the adversary's signal, not the channel gains, which we assume to be random from a known distribution. The receiver always needs the exact fading gains to decode the message, while the adversary and the transmitter may or may not know the exact values of fading gains. Therefore, we derive the capacity of the GAVFC for four cases wherein the channel gains are available at the transmitter and/or adversary as follows:
\begin{itemize}
	\item Neither the transmitter nor the adversary knows the channel gains.
	\item Only the transmitter knows the channel gains.
	\item Only the adversary knows the channel gains.
	\item Both the transmitter and the adversary knows the channel gains.
\end{itemize}
If the jammer does not know the channel gains, we show that the capacity is equal to the capacity of the corresponding fading channel with increased noise variance by the power of the jammer. If the jammer knows the fading gains, then it can choose its signal as a function of the gains, and under some power constraints it can symmetrize the channel and make the capacity zero.
Note that if the channel gains are not available at the adversary, it does not have the required channel information to symmetrize the channel. Moreover, all the results still hold if the adversary and the encoder have the channel gains causally or non-causally, except one situation. If the adversary knows the channel gains causally while the encoder knows them non-causally, then the adversary cannot symmetrize the channel since the encoder possesses some extra information that the adversary does not.

The rest of the paper is organized as follows. We describe the GAVFC model and define the various capacities in Sec. \ref{2}. We state our main theorem including the capacity of the GAVFC in all cases in Sec. \ref{3}. Before giving the proof of our main theorem, we need some auxiliary lemmas and tools which are presented in Sec. \ref{4}. Later, in sections \ref{secV}, \ref{6}, \ref{7}, and \ref{8}, we provide the converse and achievability proofs of each of the main results in Theorem \ref{Thrm}. Finally in the Appendix, we provide a brief proof for the auxiliary results.

\emph{Notation:} We use bold letters to indicate $n$-length vectors. We employ $\langle \cdot,\cdot \rangle $ and $\circ$ to denote inner product and Hadamard product (element-wise multiplication), respectively. We indicate the positive-part function, 2-norm and the expectation by $|\cdot|^+$, $\|\cdot\|$ and $\mathbb{E}[\cdot]$, respectively. Also, for an integer $N$, $[N]$ stands for the set $\{1,2,3,\ldots,N\}$. Notation $\mathbf{I}_n$ represents the identity matrix of size $n$. Each of $\log(\cdot)$ and $\exp(\cdot)$ functions has base 2. Moreover, $C(x)=\frac{1}{2}\log(1+x)$, and $X\sim\calN (\mu, \sigma^2)$ denotes Gaussian random variable $X$ with mean $\mu$ and variance $\sigma^2$.

\vspace{1ex}

\section{Problem Statement} \label{2}
\begin{figure}[t]
\centering
    \includegraphics[width=0.5\linewidth]{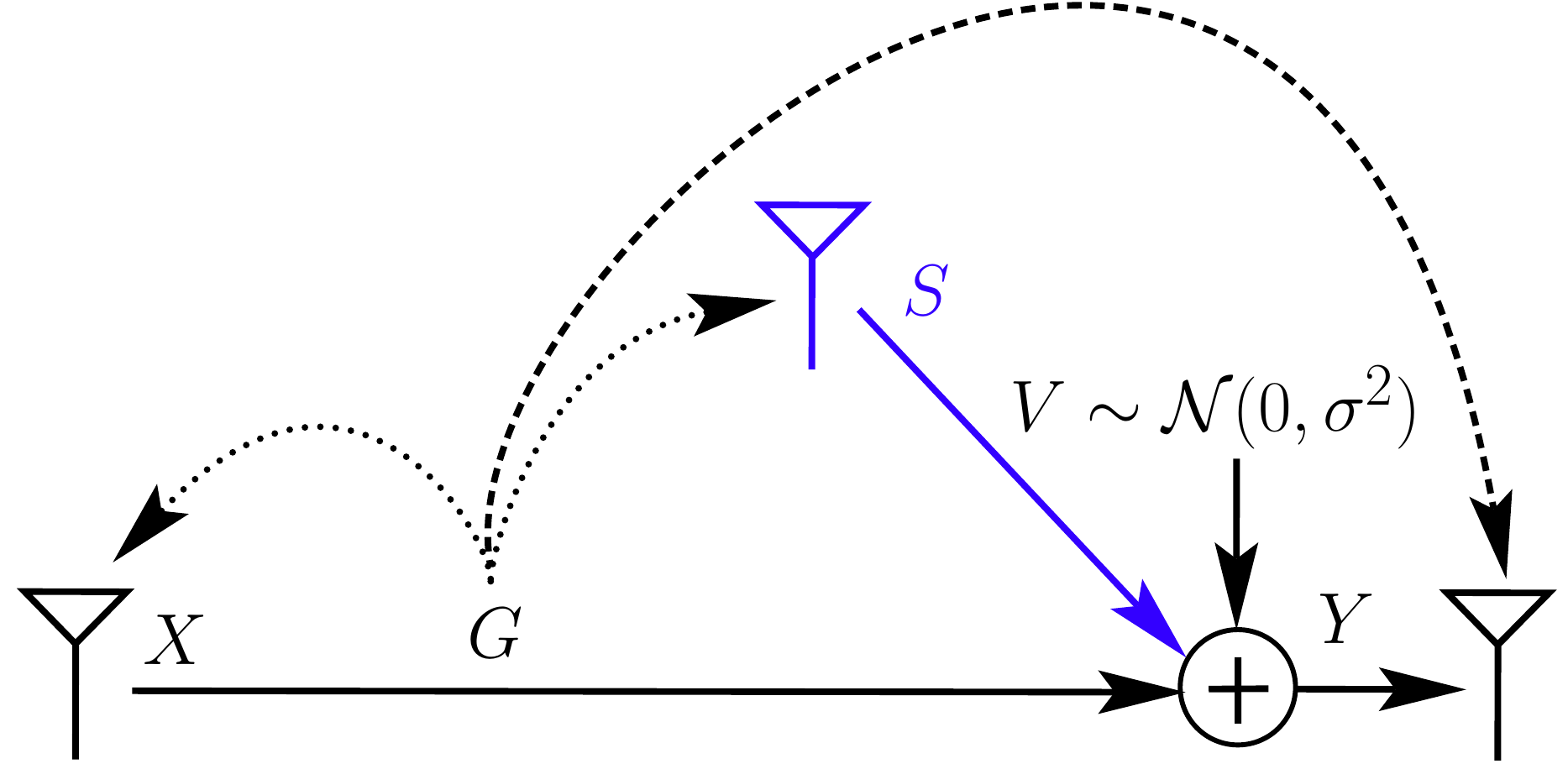}
    \caption{Gaussian Arbitrarily-Varying Fading Channel.}
    \label{fig:GAVFC}
\end{figure}

The Gaussian arbitrarily-varying fading channel (GAVFC) in Fig. \ref{fig:GAVFC} is a point-to-point fading channel with additive Gaussian noise and an intelligent adversary who does not have any information about the transmitted signal except the code. The received signal is given by 
\begin{align}\label{eq:1}
\begin{split}
\bY = \bG \circ \bx + \bs + \bV
\end{split}
\end{align} 
where $\bG$ is a random sequence of identical and independently distributed (i.i.d.) fast fading channel gains from the legitimate transmitter to the receiver drawn from continuous distribution $f_G(g)$ assumed to have positive and finite variance, $\bx$ is the $n$-length deterministic vector representing the user's signal, $\bs$ is the adversary signal chosen arbitrarily, and $\bV$ is a random $n$-length noise vector distributed as a sequence of i.i.d. zero mean Gaussian random variables with variance $\sigma^2$, independent of $\bx$, $\bG$ and $\bs$. Note that the receiver always knows the exact fading coefficients $\bg$ while the transmitter and the adversary may not know the gains, know them causally, or know them non-causally. 

Define an $\left(N,n\right)$ code for the GAVFC by a message set, an encoding function and a decoding function as follows:
\begin{itemize}
  \item Message set $\mathcal{M}=[N]$,
  \item Encoding function (one of the following)
	\begin{itemize}
	\item (No knowledge) $\bx(m):\mathcal{M}\to \mathbb{R}^n$ where $\bx = (x_1,\ldots,x_n)$,
	\item (Causal) $x_i(m,\bg^i)\!\!:\!\mathcal{M}\times \mathbb{R}^i \to \mathbb{R}$	where $\bg^i\!=\!(g_1,\ldots,g_i)$ and $\bx \!=\! (x_1,\ldots,x_n)$ for $i\in[n]$,
	\item (Non-causal) $x_i(m,\bg)\!\!:\!\mathcal{M}\times \mathbb{R}^n\!\to\! \mathbb{R}$ where $\bg\!=\!(g_1,\ldots,g_n)$ and $\bx \!=\! (x_1,\ldots,x_n)$~for~$i\!~\in~[n]$,
	\end{itemize}
	
	\item Decoding function $\Theta(\by,\bg):\mathbb{R}^n\times \mathbb{R}^n\to\mathcal{M}$,
\end{itemize}
where the rate of the code is $R = \frac{1}{n}\log (N)$. The message $m$ is drawn uniformly from the set $\mathcal{M}$. If the encoder does not know the channel gains, it maps the message to $\bx(m)\in \mathbb{R}^n$. If the encoder knows the channel gains causally, then it maps the message to $x_i(m,\bg^i)\in \mathbb{R}$, and if the encoder knows the channel gains non-causally, then it maps the message to $x_i(m,\bg)\in \mathbb{R}$ where $\bx = (x_1,\ldots, x_n)$. Given channel gains $\bg$ at the receiver, the signal $\by$ is decoded by function $\Theta(\by,\bg)$ to the message $\hat{m}$. Moreover, we assume that if the channel gains are available at the transmitter then the transmitter's signal satisfies the expected power constraints $\bbE \left[\|\bX(m,\bG)\|^2\right]\leq nP$ for any message $m\in \mathcal{M}$. Otherwise, the power constraint is $\|\bx(m)\|^2\leq nP$. The same definition applies to the adversary's signal power constraint, i.e. if the adversary knows the channel gains, the constraint is $\bbE \left[\|\bS(\bG)\|^2\right]\leq n\Lambda$; otherwise, it is $\|\bs\|^2\leq n\Lambda$. The three parameters $P$, $\Lambda$, and $\sigma^2$ as well as the distribution of fading gains $f_G(g)$ are known to all parties.

The probability of error $e(\bs,m)$ for the message $m \in \mathcal{M}$ in the presence of adversary signal $\bs\in \mathbb{R}^n$ is now given by the probability that $\hat{m}\neq m$. Thus, the average probability of error for a specific $\bs\in \mathbb{R}^n$ is
\begin{align}
	\bar{e}(\bs) = \frac{1}{N}\sum_{m=1}^N e(\bs,m).
\end{align}
If the adversary knows the channel gains non-causally then his signal is denoted by $s_i(\bg)$ for $i\in [n]$. Alternatively, if the adversary knows the gains causally, then the adversary's action is given by functions $s_i(\bg^i)$ for $i\in [n]$ where $\bs = (s_1,\cdots,s_n)$ and $\bg^i = (g_1,\cdots,g_i)$. Therefore, the average probability of error for this specific $\bs(G)\in \mathbb{R}^n$ is
\begin{align}
\bar{e} (\bs(\cdot)) = \frac{1}{N}\sum_{m=1}^N \bbE e(\bs (\bG),m).
\end{align}
Finally, the overall probability of error $P_e^{(n)}$ is maximized over all possible choices of jammers' sequences $\bs$ which satisfy either $\bbE \left[\|\bS\|^2\right]\leq n\Lambda$ or $\|\bs\|^2\leq n\Lambda$. Rate $R$ is \emph{achievable} if there exists a sequence of $\left(2^{nR},n\right)$ codes where $\underset{n \rightarrow \infty}{\lim}P_e^{(n)} = 0$. The capacity is the supremum of all achievable rates. We denote the capacity of the GAVFC as $C_{\alpha,\beta}$ where $\alpha$ denotes the transmitter's knowledge, and $\beta$ denotes the adversary's knowledge; $\alpha$ and $\beta$ can be U, C, or N depending on whether the transmitter or adversary does not know the gains (U $=$ unknown), knows the gains causally (C), or knows the gains non-causally (N). For example, $C_{\text{U,N}}$ is the capacity where the transmitter does not know the gains and the adversary knows the gains non-causally.

\section{Main Results}\label{3}
We present our results for the capacity of GAVFC whether the fading channel gains $\bG$ are available causally or non-causally at the encoder and/or the adversary (the decoder always knows the gains) in the following theorems. 

\begin{theorem}\label{Thrm} The capacities of the GAVFC are given by 
\begin{align}
	&C_{\text{U,U}} = \bbE_G \left[C\left(\frac{G^2P}{\Lambda+\sigma^2}\right)\right],\label{C1Capacity}\\
&C_{\text{N,U}} = C_{\text{C,U}} = 
	\underset{\substack{\varphi(g):\bbE \varphi(G)\leq P}}{\max}\bbE_G \left[C\left(\frac{G^2\varphi(G)}{\Lambda+\sigma^2}\right)\right],\label{C2Capacity}\\
&C_{\text{U,N}} = C_{\text{U,C}} =\begin{cases}
	\underset{\psi(g):\bbE \psi(G)\leq \Lambda}{\min}\bbE_G \left[C\left(\frac{G^2P}{\psi(G)+\sigma^2}\right)\right],\! & \bbE G^2P\!>\!\Lambda\\
	0,\! & \bbE G^2P\!\le\!\Lambda
\end{cases}\label{C3Capacity}\\
	&C_{\text{N,N}} =\! C_{\text{C,C}}\! =\! C_{\text{C,N}} =\nonumber\\
	&\begin{cases}
	\underset{\substack{\varphi(g):\bbE \varphi(G)\leq P,\\ \bbE G^2\varphi(G)\geq\Lambda}}{\max} \ \  \underset{\psi(g):\bbE \psi(G)\leq \Lambda}{\min}\bbE_G \left[C\!\left(\frac{G^2\varphi(G)}{\psi(G)+\sigma^2}\right)\!\right], & \text{if}\underset{\varphi(g):\bbE \varphi(G)\leq P}{\max}\bbE G^2\varphi(G)>\Lambda \\
	0,  & \text{if} \underset{\varphi(g):\bbE \varphi(G)\leq P}{\max}\bbE G^2\varphi(G)\le \Lambda
	\end{cases}\label{C4Capacity}\\
	&C_{\text{N,C}} = 
	\underset{\substack{\varphi(g):\bbE \varphi(G)\leq P}}{\max} \ \  \underset{\psi(g):\bbE \psi(G)\leq \Lambda}{\min}\bbE_G \left[C\!\left(\frac{G^2\varphi(G)}{\psi(G)+\sigma^2}\right)\!\right].\label{C42Capacity}
\end{align}
\end{theorem}

Note that when the encoder knows the gains \eqref{C2Capacity}, \eqref{C4Capacity}, and \eqref{C42Capacity}, the capacity expression includes a maximization of the input power as a function $\varphi(\cdot)$ of the gain, similar to the result in \cite{Varaiya}. Similarly, when the jammer knows the gains \eqref{C3Capacity}--\eqref{C42Capacity}, the capacity expression includes a minimization that represents the jammer's choice of noise power as a function $\psi(\cdot)$ of the gain. Moreover, when the jammer knows the gains, with enough power it can symmetrize the channel by mimicking the legitimate signal, thus reducing the capacity to zero. However, in \eqref{C42Capacity} we have assumed that the adversary knows the gains causally and the encoder and the decoder know the gains non-causally. Thus, the encoder and decoder effectively share a secret (the channel gains at the end of the block) unknown to the adversary, so the adversary cannot symmetrize the channel. It is also worth mentioning that for the other cases (except \eqref{C42Capacity}) our proof works exactly the same whether the transmitter and/or the adversary know the gain sequence causally, non-causally, or even memorylessly (i.e., at time $i$, you only know the gain value at time $i$). 

While we have stated the theorem by writing the capacities in terms of optimization over the $\varphi$ and/or $\psi$ functions, these expressions can be computed by solving for the optimizing functions. In particular, the optimum value of $\varphi^*(g)$ in \eqref{C2Capacity} is $\left|\lambda-\frac{\Lambda+\sigma^2}{g^2}\right|^+$ where $\lambda$ is obtained by $\bbE [\varphi^*(G)]=P$, and 
the optimum value of $\psi^*(\cdot)$ in \eqref{C3Capacity} is a function of gain $g$ as follows
\begin{align}
\psi^*(g) = \left|\frac{-2\sigma^2-g^2P+\sqrt{g^4P^2+\frac{2g^2P}{\lambda}}}{2}\right|^+,\label{fi_g3}
\end{align}
and $\lambda$ is obtained by solving $\bbE \psi^*(G) = \Lambda$. Moreover, the optimum values of $\varphi^*(g)$ and $\psi^*(g)$ in \eqref{C4Capacity} are 
\begin{align}
\varphi^*(g) &= \left| \frac{1}{2(\lambda_1-g^2\lambda_3) \left(1+\frac{\lambda_1-g^2\lambda_3}{g^2\lambda_2}\right)}\right|^+\\
\psi^*(g) &= \left|\frac{g^2}{2g^2\lambda_2+2(\lambda_1-g^2\lambda_3)}-\sigma^2\right|^+,
\end{align}
where $\lambda_1$, $\lambda_2$ and $\lambda_3$ are found by solving $\bbE \varphi(G)= P$, $\bbE \psi(G)= \Lambda$ and $\bbE G^2\varphi(G)=\Lambda$, respectively.
Finally, the optimum values of $\varphi^*(g)$ and $\psi^*(g)$ in \eqref{C42Capacity} are
\begin{align}
\varphi^*(g) &= \left| \frac{1}{2\lambda_1 \left(1+\frac{\lambda_1}{g^2\lambda_2}\right)}\right|^+\\
\psi^*(g) &= \left|\frac{g^2}{2g^2\lambda_2+2\lambda_1}-\sigma^2\right|^+,
\end{align}
where $\lambda_1$ and $\lambda_2$ can be obtained by solving $\bbE \varphi(G)= P$ and $\bbE \psi(G)= \Lambda$, respectively.

\begin{figure}[t]
\centering
    \includegraphics[width=0.7\linewidth]{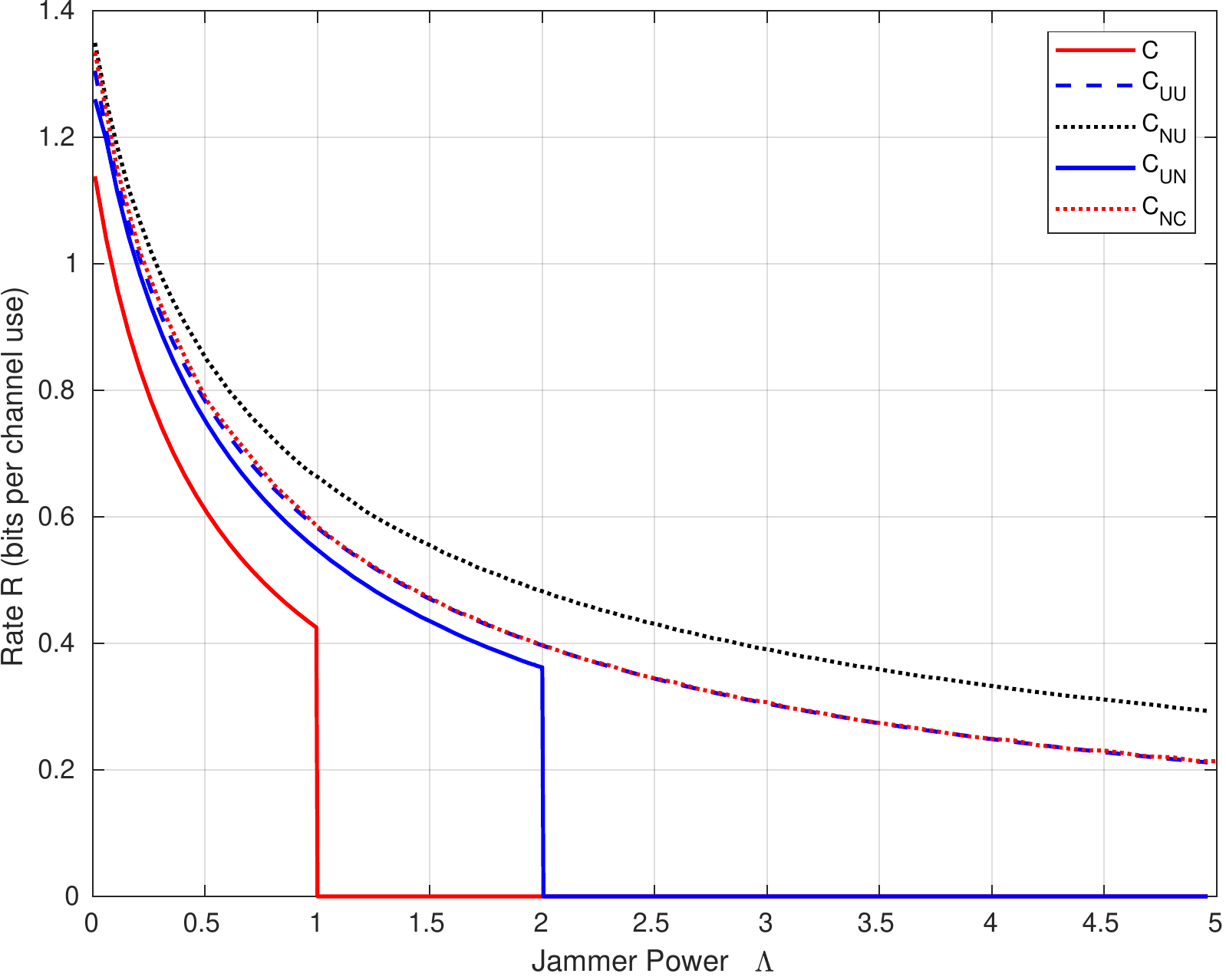}
    \vspace{-1em}\caption{GAVFC capacities for $P=1, \sigma^2=0.25, 0<\Lambda<5$ with Rayleigh fading. $C$ is the capacity of standard Gaussian channel without fading.}
    \label{fig:GAVFC_Simu}
\end{figure}

In the Fig. \ref{fig:GAVFC_Simu}, the capacity of GAVFC with Rayleigh fading is shown for $P=1, \sigma^2=0.25, 0<\Lambda<5$ and the Rayleigh distribution scale parameter $\sigma_R = 1$ whether the channel gains are available at the encoder and/or the adversary. However, $C$ is the capacity of Gaussian arbitrary-varying channel with no fading. Note that when the transmitter's knowledge increases, the capacity increases, whereas when the adversary's knowledge increases, the capacity decreases. On the other hand, the knowledge of adversary about the channel gains may decrease the capacity, and in this case if the adversary's power exceeds $2$, the capacity will be zero by the symmetrizability.

The proofs of the different capacity variants all follow a similar pattern, so we have attempted to reduce the redundancy in our presentation. We provide essentially one converse proof for each capacity expression in the theorem:
\begin{enumerate}
\item Sec. \ref{subsecV-A}: Converse for $C_{U,U}$.
\item Sec. \ref{subsecVI-A}: Converse for $C_{N,U}$. This also bounds $C_{C,U}$ which is trivially upper bounded by $C_{N,U}$.
\item Sec. \ref{subsecVII-A}: Converse for $C_{U,C}$. This also bounds $C_{U,N}$ which is trivially upper bounded by $C_{U,C}$.
\item Sec. \ref{subsecVIII-A}: Converse for $C_{C,C}$. This also bounds $C_{C,N}$ which is trivially upper bounded by $C_{C,C}$. Essentially the same proof also works for $C_{N,N}$. The case $C_{N,C}$ requires a different proof also covered in this section.
\end{enumerate}
We provide essentially three achievability proofs:
\begin{enumerate}
\item Sec. \ref{subsecVI-B}: Achievability for $C_{C,U}$. This also bounds the $C_{N,U}$, which is trivially lower bounded by $C_{C,N}$. It also provides a bound for $C_{U,U}$, because the same proof works assuming $\varphi(g)=P$ (i.e., the encoder's power is independent of the channel gain).
\item Sec. \ref{subsecVIII-B}: Achievability for $C_{C,N}$. This also bounds $C_{N,N}$ and $C_{C,C}$, which are trivially lower bounded by $C_{C,N}$. It also provides a bound for $C_{U,N}$ and $C_{U,C}$, because the same proof works again assuming $\varphi(g)=P$.
\item Sec. \ref{subsecVIII-C}: Achievability for $C_{N,C}$. This case is different from all others in that there is effectively a shared secret between encoder and decoder.
\end{enumerate}

\section{Auxiliary Results and Tools}\label{4}

Before proceeding to the proofs, we first define the typical set for continuous random variables $X_1,\ldots,X_k$ with probability density function $f_{X_1,\ldots,X_k}(x_1,\ldots,x_k)$ as follows:
\begin{multline}
\T(X_1,\ldots,X_k)=
\bigg\{(x_1,\ldots,x_k)\!:
\left|-\frac{1}{n}\log f_{X_A}(x_{A})
-h(X_A)\right| \leq \eps  
\text{ for all }A\!\subset\![k]\!\bigg\}\label{TypicalSet}
\end{multline}
where $h(X_A)$ is the differential entropy of $(X_i:i\in A)$. Next, we define the typical set for continuous random variables $X_1,\ldots,X_k$ with probability density function $f_{X_1,\ldots,X_k}(x_1,\ldots,x_k)$ and a discrete random variable $\tilde{G}$ with probability mass function $P_{\tilde{G}(\tilde{g})}$ as follows:
\begin{multline}
\T(X_1,\ldots,X_k,\tilde{G})=
\bigg\{(x_1,\ldots,x_k,\hat{g})\!:
\left|-\frac{1}{n}\log P_{\tilde{G}}(\tilde{g})-H(\tilde{G})\right| \leq \eps ,\\
\left|-\frac{1}{n}\log f_{X_A}(x_{A})
-h(X_A)\right| \leq \eps,
\left|-\frac{1}{n}\log f_{X_A
|\tilde{G}}(x_{A}|\tilde{g})
-h(X_A|\tilde{G})\right| \leq \eps ,   
\text{ for all }A\!\subset\![k]\!\bigg\},\label{TypicalSet}
\end{multline}
where $H(\tilde{G})$ and $h(X_A|\tilde{G})$ denote the entropy of $G$ and the conditional differential entropy of $X_A$ given $\tilde{G}$.

 Throughout the achievability proofs, we will utilize several lemmas including the \emph{joint typicality lemma} and \emph{conditional typicality lemma} for Gaussian random variables given in \cite{FatemehISIT2018}. In addition, we will need the following two lemmas; they show that with high probability a Gaussian codebook satisfies several desirable properties. The proof is given in the Appendix.

\begin{lemma}\label{lem5_0} Fix $\epsilon'>0$. There exists $\gamma>0$ such that the following holds. Let $\mathbf{X}(m)$ for $m\in[N]$, $N=2^{nR}$ be a zero mean Gaussian codebook with variance $1-\gamma$. Let $G$ be drawn from probability density function $f_G(g)$. With probability approaching 1 as $n\to\infty$, for any $\bs,\bg$ where $\|\bs\|^2\le n\Lambda$,
there exists a function $\delta(\eps')>0$ such that
\begin{align}
\frac{1}{N} \left|\left\{m: (\bx(m),\bs,\bg)\notin \hspace{-1.6em} \bigcup_{\substack{X\text{ independent of } (S,G): \\EX^2=1, ES^2\le\Lambda}}\hspace{-2.1em}\mathcal{T}^{(n)}_{\eps'}(X,S,G)\right\}\right|\le \exp(-n\delta(\eps')),\label{eq:two_codebooks_c0}
\end{align}
where the union is over zero mean conditionally Gaussian random vectors $(X,S)$ given $G$.
\end{lemma}

\begin{lemma}\label{lem5} Fix $\epsilon>0$. There exists $\gamma>0$ such that the following holds. Let $\mathbf{X}(m)$ for $m\in[N]$, $N=2^{nR}$ be a zero mean Gaussian codebook with variance $1-\gamma$. Let $G$ be drawn from probability density function $f_G(g)$. With probability approaching 1 as $n\to\infty$, for any 
\bi 
\m zero-mean conditionally Gaussian random vector $(X,X',S)$ given $G$ where $\bbE X^2=\bbE X'^2=1$ and $\bbE S^2\le \Lambda$,
\m $\bx,\bs,\bg$ where $\|\bs\|^2\le n\Lambda$,
\ei 
there exists a function $\delta(\eps)>0$ such that
\begin{align}
&\bbP \left\{\big|\big\{(\bx(m'),\bs,\bG)\!\in\!\T\!(X',S,G)\text{ for some }m'\big\}\big|\right\} \le 2 \exp\{\!-n\delta(\eps)/2\},\nonumber \\& \hspace{20em}\text{if }I(G;X'S)\!\ge\! |R\!-\! I(X';S)|^+\!\!+\! \delta(\eps),\label{eq:Hughes_c4}\\
&\big|\big\{m':(\bx(m'),\bs)\in\T(X',S)\big\}\big| \le \exp\big\{n\big[|R-I(X';S)|^++\delta(\eps)\big]\big\},\label{eq:Hughes_c5}
\\
&\big|\big\{m':(\bx,\bx(m'),\bs,\bg)\in\T(X,X',S,G)\big\}\big|\le \exp\big\{n\big[|R-I(X';XSG)|^++\delta(\eps)\big]\big\},\label{eq:22Huz_c5}\\
&\frac{1}{N}\big|\big\{\! m\!:\!(\bx(m),\bx(m'),\bs,\bg)\!\in\!\T\text{ for some }m'\!\ne\! m\!\big\}\big|\le \!2\exp\{\!-n\delta(\eps)/2\},\nonumber\\
&\hspace{19em}\text{if }I(X;X'SG)\!\ge\! |R\!-\! I(X';SG)|^+\!\!+\! \delta(\eps).\label{eq:21Huz_c4}
\end{align}
\end{lemma}

\section{Capacity Proof with Gains Available at Decoder}\label{secV}

\subsection{Converse Proof}\label{subsecV-A}
We initially assume that for any arbitrary adversary strategy there is a sequence of $(2^{nR},n)$ codes with vanishing probability of error. The adversary can generate a Gaussian sequence with variance $\Lambda-\gamma$ for any $\gamma>0$; if this sequence has power less than $\Lambda$, it is transmitted, otherwise, the adversary sends the all-zero sequence. Note that the power of this Gaussian sequence exceeds $\Lambda$ only with small probability by the law of large numbers. With this choice of adversary, the channel corresponds to a standard Gaussian fading channel with the noise variance $\Lambda+\sigma^2-\gamma$ where the channel gains are available only at the decoder. Therefore, using capacity of a non-adversarial Gaussian fading channel \cite{ElGamal} for arbitrarily small $\gamma$, we may upper bound the capacity by
\be
C\le \bbE_G \left[C\left(\frac{G^2P}{\Lambda+\sigma^2}\right)\right].
\ee

\subsection{Achievability Proof}\label{subsecV-B}

The achievability proof of this case can be counted as a special case of $C_{N,U}$ in Sec. \ref{subsecVI-B} where both encoder and decoder know the channel gains. However, in this case since the encoder does not know the channel gains, we do not have any $\varphi(g)$ function at the encoder. In other words, the achievability proof for this case is identical to that in Sec. \ref{subsecVI-B} with $\varphi(g)=P$.

\section{Capacity Proof with Gains Available at Encoder and Decoder}\label{6}

\subsection{Converse Proof}\label{subsecVI-A}

As in the previous case, the adversary can simply send Gaussian send noise with variance $\Lambda-\gamma$. By the law of large numbers, the resulting channel is equivalent to a standard Gaussian fading channel with the knowledge of gains at both encoder and decoder and noise variance $\Lambda+\sigma^2-\gamma$ with high probability. Thus, since $\gamma$ can be chosen arbitrarily small, from the capacity of a non-adversarial Gaussian fading channel  \cite{Varaiya}, we have
\be
C\le \underset{\varphi(g):\bbE \varphi(G)\leq P}{\max}   \bbE_G \left[C\left(\frac{G^2\varphi(G)}{\Lambda+\sigma^2}\right)\right].
\ee

\subsection{Achievability Proof}\label{subsecVI-B}

For simplicity we assume $P=1$. Suppose any arbitrary function $\varphi(G)$ that satisfies $\bbE \varphi(G)\leq 1$ and $\var(G\sqrt{\varphi(G)})>0$. We further assume that $G^2 \varphi(G)$ has a positive variance. Note that this is only a concern if the optimum $\varphi^*(G)=\frac{c}{G^2}$; in this case, we can instead take $\varphi(G)=\frac{c}{(G-d)^2}$ where $c,d$ are two positive constants and $d$ can be chosen arbitrarily small. Let
\begin{align}
	R<\bbE_G \left[C\left(\frac{G^2\varphi(G)}{\Lambda+\sigma^2}\right)\right]. \label{eq:R_assumption2}
	\end{align}
We now propose a $(2^{nR},n)$ code sequence, and prove that using this code the probability of error tends to zero as $n\to \infty$.

\emph{Codebook generation:} Fix $\eps>\eps'>\gamma>0$. We generate $2^{nR}$ i.i.d zero mean Gaussian sequences $\bX(m)$ with variance $(1-\gamma)$ for each $m \in [2^{nR}]$. By Lemma~\ref{lem5_0} and Lemma \ref{lem5}, we assume that the deterministic codebook satisfies \eqref{eq:two_codebooks_c0}--\eqref{eq:21Huz_c4}.

\emph{Encoding:} Since the transmitter knows the channel gains, it sends $\sqrt{\varphi(\bg)}\circ \bx(m)$ (at time $i$ signal $\sqrt{\varphi(g_i)}x_i(m)$ is sent) if its power is less than $1$, otherwise it sends zero. 

\emph{Decoding:} Given $\by$, let $\mathscr{S}$ be the set of messages $\hat{m}$ such that $(\bx(\hat{m}),\bg, \by)\in \calT_\eps^{(n)}(X',G,Y)$ for some random variables $X'\sim \calN(0,1)$, $G\sim f_G(g)$ and zero mean Gaussian $Y-G\sqrt{\varphi(G)}X'$ where $(X',G,Y-G\sqrt{\varphi(G)}X')$ are mutually independent.

Now, we define the decoding function as
\begin{align}
	\Theta(\mathbf{y},\bg) &= \argmin_{\hat{m} \in  \mathscr{S}}
 \left\| \by - \bg \circ \sqrt{\varphi(\bg)}\circ \bx(\hat{m}) \right\|^2.
\end{align}

\emph{Analysis of the probability of error:}
Suppose the true message sent by the legitimate user is message $M$ with the power constraint $\|\bx(M)\|^2\le n(1-\gamma)$. Then, the overall probability of error is upper bounded by $P_e^{(n)}\leq P_0+P_1$ where 
\begin{align}
P_0&=\bbP\left\{M \notin\mathscr{S} \right\},\label{firstPe}\\
P_1&=\bbP\bigg\{\left\| \bY - \bG\circ\sqrt{\varphi(\bG)}\circ\bx(\hat{m}) \right\|^2
\leq \left\| \bs+\bV \right\|^2 \text{ for some } \hat{m}\in \mathscr{S}\setminus \{M\} \bigg\}.\label{secondPe}
\end{align}
Consider any state sequence $\mathbf{s}$. By \eqref{eq:two_codebooks_c0}, with high probability $(\mathbf{x}(M),\mathbf{s},\bG)\in \mathcal{T}_{\eps'}^{(n)}(X,S,G)$ where $(X,S,G)$ are independent, and $\bbE X^2=1,\allowbreak \bbE S^2 \le \Lambda$. By the conditional typicality lemma, for every $\eps>\eps'$ with high probability $(\bx(M),\bs,\bG,\bV)\in \T(X,S,G,V)$ where $(X,S,G,V)$ are mutually independent, and $\bbE V^2=\sigma^2$. 
Thus, according to the definition of $\mathscr{S}$, with high probability $M\in \mathscr{S}$ and $P_0$ tends to zero as $n\to \infty$.

Define the shorthand $\vec{X}=(XX'SGV)$. Let $\mathcal{V}$ be a finite $\eps$-dense subset in the set of all distributions of random vectors $\vec{X}$ that are determined by $f_G(g)$ and jointly zero mean Gaussian vector $(XX'SV)$ independent of $G$ with bounded covariances at most $(1,1,\Lambda,\sigma^2)$. Note that because the distribution of $f_G(g)$ is fixed, the overall distribution of $\vec{X}$ can be determined by the covariance matrix of $(XX'SV)$, so $\mathcal{V}$ only needs to cover a compact set. Now, we may upper bound $P_1$ by
\begin{equation}
\sum_{\vec{X}\in\mathcal{V}} \frac{1}{N} \sum_{m=1}^N \bbE_{G}[e_{\vec{X}}(m,\mathbf{s},\bG)]
\end{equation}
where 
\begin{multline}\label{eq:eX1_def2}
e_{\vec{X}}(m,\mathbf{s},\bg)
=\mathbb{P}\bigg\{ (\bx(m),\bx(\hat{m}),\bs,\bg,\bV)\in\calT_\eps^{(n)}(\vec{X}),\\
\|\bg \circ \sqrt{\varphi(\bg)} \circ \bx(m)+\bs+\bV-\bg \circ \sqrt{\varphi(\bg)}\circ \bx(\hat{m})\|^2 \le \|\bs+\bV\|^2 \text{ for some }\hat{m}\in\mathscr{S}\setminus \{m\}\bigg\}.
\end{multline}
We will show that $\frac{1}{N} \sum_{m=1}^N e_{\vec{X}}(m,\mathbf{s},\bg)\to 0$ for all vectors $\bg$ and all vectors $(XX'SV)$ which are Gaussian given $G$ (whether or not they are in $\calV$). Let $Z =G\sqrt{\varphi(G)}X+S+V-G\sqrt{\varphi(G)}X'$. We may restrict ourselves to $\vec{X}$ where
\begin{gather}
(X,S,G,V)\text{ are mutually independent},\label{XSV_indep}\\
(X,X',S,V)\text{ are zero mean Gaussian},\label{zero_mean}\\
\bbE X^2=\bbE X'^2=1,\quad
\bbE V^2=\sigma^2,\quad \bbE S^2\le\Lambda,\label{signalpower}\\
\left(X',G,Z\right)\!\text{ are independent},\label{XZ1_indep}\\
\bbE \left[Z^2\right] \le \Lambda+\sigma^2,\label{Zles}
\end{gather}
where \eqref{XSV_indep} holds since the input $X$, adversary $S$, fading gains $G$ and noise $V$ are all generated independently, \eqref{zero_mean}--\eqref{signalpower} follows from $m,\hat{m}\in\mathscr{S}$, and $\vec{X}\in\mathcal{V}$, \eqref{XZ1_indep} holds since we have $(X',G,Y-GX')$ are mutually independent using $\bx(\hat{m})\in\mathscr{S}$, and \eqref{Zles} corresponds to $\bbE \left[\left(Y-G\sqrt{\varphi(G)}X'\right)^2\right]$ which is less than $\Lambda+\sigma^2$ from \eqref{eq:eX1_def2}. 

Observe that if $I(X,V,G;X',S)=0$, then we would have
\begin{align}
0&=\bbE [X'Z]\label{G_indep2}\\
&=\bbE [X'(G\sqrt{\varphi(G)}X+S+V-G\sqrt{\varphi(G)}X')]\\
&= \bbE [X'(S-G\sqrt{\varphi(G)}X')]\label{28_2}\\
&= \bbE [X'S] - \bbE G\sqrt{\varphi(G)},\label{contr3}
\end{align}
where \eqref{G_indep2} follows from \eqref{XZ1_indep}, \eqref{28_2} holds because $(X',G,X,V)$ are all mutually independent by the assumption $I(X,V,G;X',S)=0$ and \eqref{XSV_indep}, and the last equality holds since $X'$ is independent of $G$ and because $\bbE [X'^2]=1$. Therefore, $\bbE [X'S] = \bbE G\sqrt{\varphi(G)}$. 

Moreover, from \eqref{eq:eX1_def2} we have
\begin{align}
\bbE (S+V)^2&\geq \bbE(G\sqrt{\varphi(G)}X\!+\!S\!+\!V\!-\!G\sqrt{\varphi(G)}X')^2
\\&=\bbE G^2\varphi(G)(X-X')^2+2\bbE G\sqrt{\varphi(G)}(X-X')(S+V)+\bbE(S+V)^2
\\&=\bbE G^2\varphi(G)\bbE X^2+\bbE G^2\varphi(G)\bbE X'^2-2\bbE G\sqrt{\varphi(G)}X'S+\bbE(S+V)^2\label{lotwork}
\\&=2\bbE G^2\varphi(G)-2\bbE G\sqrt{\varphi(G)}\bbE X'S+\bbE(S+V)^2,\label{cancel1}
\end{align}
where \eqref{lotwork} holds because $\bbE X = \bbE X' = \bbE V = 0$, $(X,X',G)$ are mutually independent, $(X,S,V)$ are mutually independent, and $(X',V)$ are independent by \eqref{XSV_indep}, \eqref{zero_mean} and the assumption $I(X,V,G;X',S)=0$. Canceling $\bbE(S+V)^2$ from both sides of \eqref{cancel1} gives us 
\begin{align}
	\bbE G^2\varphi(G)-\bbE G\sqrt{\varphi(G)}\bbE X'S\leq 0.\label{contra2}
\end{align}
Now, if we apply the result from \eqref{contr3} to \eqref{contra2}, we get 
\begin{align}
		\bbE G^2\varphi(G)-\bbE G\sqrt{\varphi(G)}\bbE X'S&=\bbE G^2\varphi(G)-\bbE G\sqrt{\varphi(G)}\bbE G\sqrt{\varphi(G)}\\
		&=\bbE G^2\varphi(G)-\bbE^2 G\sqrt{\varphi(G)}\\
		&=\var{G\sqrt{\varphi(G)}}\\
		&\leq 0.
\end{align}
which is a contradiction since we assume $\var{(G\sqrt{\varphi(G)})}$ is always positive. Thus, there exists an $\eta>0$ such that
\be\label{eq:eta_bound12}
\eta\le I(XVG;X'S).
\ee

Also, by \eqref{eq:21Huz_c4}, we may restrict ourselves to distributions where
\be\label{eq:R1R212}
I(X;X'SG)< |R-I(X';SG)|^++\delta(\eps)
\ee
and 
\be\label{eq:RI(X'S)2}
I(G;X'S)< |R-I(X';S)|^++\delta(\eps).
\ee
Note that $I(X;X'SG)=I(X;X'|SG)$. We also have the upper bound
\begin{align}
e_{\vec X}(m,\bs,\bg)&\leq\sum_{\hat{m}:(\bx(m),\bx(\hat{m}),\bs,\bg)\in\T(X,X',S,G)}
\hspace{-2em}\bbP\left\{\!(\bx(m),\bx(\hat{m}),\bs,\bg,\bV)\!\in\! \T (X,X',S,G,V)\!\right\}
\\& \leq \exp\big\{n\big[|R\!-\! I(X';XSG)|^+\!\!-I(V;X'|XSG)+\delta(\eps)\big]\label{e(s,i)_I2}
\end{align}
where \eqref{e(s,i)_I2} follows from  $I(V;XSG)=0$, \eqref{eq:22Huz_c5} and the joint typicality lemma.

Now, let us consider three cases as follows:

Case (a): $R<I(X';S)$ that implies $R<I(X';XSG)$.
From \eqref{e(s,i)_I2}, for any $m, \bs,\bg$ 
\begin{align}
e_{\vec X}(m,\bs,\bg)&\le \exp\left\{-n\left( I(V;X'|XSG)-\delta(\eps)\right)\right\}\label{eq:two_codebook_ebound11}\\
&= \exp\{- n(I(XV;X'|SG)-I(X;X'|SG)-I(XV;S|G)-\delta(\eps))\}\\
&= \exp\{- n(I(XV;X'S|G)\!-\! I(X;X'|SG)- \delta(\eps))\}\\
&= \exp\{- n(I(XVG;X'S)-I(G;X'S)-I(X;X'|SG)-\delta(\eps))\}\\
&\le \exp\{-n(\eta-\delta(\eps)-\delta'(\eps))\}\label{casearesult2}
\end{align}
where \eqref{casearesult2} follows from \eqref{eq:eta_bound12}, \eqref{eq:R1R212} and \eqref{eq:RI(X'S)2}. Therefore, $e_{\vec X}(m,\bs,\bg)$ vanishes exponentially fast if $\delta(\eps)$ is sufficiently small.

Case (b): $I(X';S)\leq R$. Since $R\geq I(X';S)$ and $I(G;S)=0$, from \eqref{eq:RI(X'S)2} we have
\begin{align}
R &> I(G;X'S)+I(X';S)-\delta(\eps)\\
& =I(G;S)+I(G;X'|S)+I(X';S)-\delta(\eps)\\
& =I(X';SG)-\delta(\eps).
\end{align}
Using this result in \eqref{eq:R1R212}, we have 
\begin{align}
I(X;X'SG)<R-I(X';SG)+\delta(\eps)+\delta(\eps).
\end{align}
Therefore,
\begin{align}
R &>I(X;X'SG)+I(X';SG)-2\delta(\eps)\\
&\geq I(X';XSG)-2\delta(\eps).\label{neweq}
\end{align}
Now, from \eqref{e(s,i)_I2}, we have for any $m, \bs,\bg$ 
\begin{align}
e_{\vec X}(m,\bs,\bg)&\leq \exp\big\{n\big[|R\!-\! I(X';XSG)|^+\!\!-I(V;X'|XSG)+\delta(\eps)\big]\\
&\leq \exp\big\{n\big[R\!- I(X';XSG)+2\delta(\eps)-\! I(V;X'|XSG)\!+\!\delta(\eps)\big]\label{abslt2}\\
& = \!\exp (n[R-I(X';XSGV)+3\delta(\eps)])\label{result222}
\end{align}
where \eqref{abslt2} follows from \eqref{neweq}.
We now lower bound $I(X';XSVG)$ as follows: 
\begin{align}
I(X';XSVG)&=I(X';XSV|G)+I(X';G)\\
 & \geq I(X';G\sqrt{\varphi(G)}X\!+\!S\!+\!V|G)\\
& = I(X';Z+G\sqrt{\varphi(G)}X'|G)\\
& = h(Z+G\sqrt{\varphi(G)}X'|G)-h(Z+G\sqrt{\varphi(G)}X'|G,X')\\
& = \bbE \bigg[\frac{1}{2} \log 2\pi e\left(G^2\varphi(G)+\bbE [Z^2|G]\right)-\frac{1}{2} \log 2\pi e \bbE [Z^2|G] \bigg] \\
& = \bbE \left[C\left(\frac{G^2\varphi(G)}{\bbE [Z^2|G]} \right) \right] \\
& \geq \bbE \left[C\left(\frac{G^2\varphi(G)}{\Lambda+\sigma^2}\right)\right]\label{ZleLambda}
\end{align}
where \eqref{ZleLambda} follows from \eqref{XZ1_indep} and \eqref{Zles}. Replacing this result in \eqref{result222}, we obtain
\be
e_{\vec X}(m,\bs,\bg)\le\exp\left\{n\left[R-\bbE \left[C\left(\frac{G^2\varphi(G)}{\Lambda+\sigma^2}\right)\right]+3\delta(\eps)\right]\right\}\label{e_1result}
\ee
meaning that $e_{\vec X}(m,\bs,\bg)$ is exponentially vanishing if $\delta(\eps)$ is sufficiently small, and \eqref{eq:R_assumption2} holds.

\section{Capacity Proof with Gains Available at Decoder and Jammer}\label{7}

\subsection{Converse Proof}\label{subsecVII-A}
Consider a sequence of $(2^{nR},n)$ codes with vanishing probability of error that must function for arbitrary jamming signals. Because we are proving the converse, we may assume the best case scenario from the legitimate user's perspective; in particular, that the adversary only knows the channel gains causally.

We begin with the case that $\Lambda\leq\bbE G^2P$. Given any function $\psi(g)$ satisfying $\bbE \psi(G) \le \Lambda$, we may obtain an upper bound by assuming that the jammer transmits a random sequence $\bS=(S_1,\cdots,S_n)$ where $S_i$ is Gaussian with mean zero and variance $\psi(G_i)$ for $i=1,\cdots,n$. Note that 
\begin{align}
\bbE [\|\bS\|^2]& = \bbE \sum_{i=1}^n S_i^2\\
&  =  \sum_{i=1}^n \bbE S_i^2\\
& =  \sum_{i=1}^n \bbE \psi(G_i)\\
& \le n\Lambda.
\end{align}
The resulting channel is equivalent to a standard Gaussian fading channel with the knowledge of gains only at the decoder and noise variance $\psi(g)+\sigma^2$. From the capacity of a non-adversarial Gaussian fading channel
\be
C\le \bbE_G \left[C\left(\frac{G^2P}{\psi(G)+\sigma^2}\right)\right].
\ee
Therefore, the capacity is also less than the minimum over all $\psi(G)$ that satisfies $\bbE \psi(G)\leq \Lambda$.
\be
C\le \min_{\psi(G):\bbE \psi(G)\leq \Lambda} \bbE_G \left[C\left(\frac{G^2P}{\psi(G)+\sigma^2}\right)\right].
\ee

For the case $\Lambda>\bbE G^2P$, we first show that the adversary has enough power to choose a codeword and send it to the channel, thereby symmetrizing the channel. Let $\tilde{M}$ be a uniformly chosen message by the adversary and $M$ be the true message sent by the legitimate transmitter. Suppose the adversary chooses $\bS=\bG \circ \bx(\tilde{M})$ then the adversary's power constraint is satisfied as follows:
\begin{align}
\bbE \left[\|\bS\|^2\right] & =\bbE \left[\|\bG \circ \bx(\tilde{M})\|^2\right] \\
& = \bbE \left[\sum_{i=1}^n G_i^2 x_i^2(\tilde{M})\right]\\
& <\sum_{i=1}^n x_i^2(\tilde{M}) \frac{\Lambda}{P}\label{last1}\\
& \le n\Lambda\label{last2}
\end{align}
where \eqref{last1} follows from the assumption $\Lambda > \bbE G^2 P$, and \eqref{last2} follows from the codebook power constraint $\|\bx^2\|\le nP$.
Given this choice of $\bS$, $\bY = \bG\circ\bx(M)+ \bG\circ \bx(\tilde{M})+\bV$. Thus, with high probability the decoder cannot decode the message since it does not know whether the true message is $M$ or $\tilde{M}$.

\subsection{Achievability Proof}\label{subsecVII-B}

The achievability proof of this case is very similar to the achievability proof of Sec.~\ref{subsecVIII-B} where the encoder, the decoder and the adversary all know the channel gains. 
Here, the transmitter does not know the channel gain so it cannot leverage this knowledge to choose its transmit power. However, the achievability proof for this case is identical to that in Sec.~\ref{subsecVIII-B} except that the transmitter's power function is constant; i.e., $\varphi(g)=1$.

\section{Capacity Proof with Gains Available at Encoder, Decoder, and Jammer}\label{8}
In this section, we first provide the converse proof for the case that the channel gains are available at the encoder, the decoder and the adversary in Sec. \ref{subsecVIII-A}. The converse proof includes all the four cases in which each of the adversary and the encoder knows the fading gains causally or non-causally. In Sec. \ref{subsecVIII-B}, we show the achievability proof of the case that the channel gains are available non-causally at the adversary and causally at the encoder. This proof also works for the two cases of channel gains being available causally at both the adversary and the encoder or non-causally at both ends. Finally, we provide the achievability proof for the last case when the channel gains are causally available at the adversary and non-causally available at the encoder in Sec. \ref{subsecVIII-C}. 

\subsection{Converse Proof}\label{subsecVIII-A}

Consider a sequence of $(2^{nR},n)$ codes with vanishing probability of error. Since in this case both the encoder and the adversary know the channel gains, we consider four cases to prove the converse whether each of them knows the fading gains causally or non-causally. 

First assume that both the encoder and adversary know the channel gains causally. Let $\varphi_i(g)=\frac{1}{N}\sum_{m=1}^N \bbE [ X_i^2(m,G^i)|G_i=g]$ and $\varphi(g)=\frac{1}{n}\sum_{i=1}^n\varphi_i(g)$ where $G^i=(G_1,\ldots,G_i)$, for $i\in[n]$. Thus, $\varphi(g)$ satisfies $\bbE \varphi(G) \leq P$ as follows:
\begin{align}
\bbE \varphi(G) & = \bbE \left[\frac{1}{n}\sum_{i=1}^n \varphi_i(G)\right] \\ 
& =\frac{1}{N}\sum_{m=1}^N\frac{1}{n}\bbE \left[\sum_{i=1}^n X_i^2(m,G^i)\right] \\
& \le P\label{powassum}
\end{align}
where \eqref{powassum} follows by the power constraint for the input signal. 

Now, similar to the previous case, where the adversary and decoder know the channel gains, we also have symmetrizability and non-symmetrizability cases, but with different conditions. We first show the symmetrizability case, that is if $\Lambda\ge \bbE G^2\varphi(G)$, then the jammer can symmetrize the channel. Suppose the adversary chooses a message $\tilde{M}$ uniformly at random and sends $S_i = G_i X_i(\tilde{M},G^i)$ where $G^i = (G_1,\cdots,G_i)$ for $i\in[n]$. Note that this selection of jamming signal is a causal function of the channel gains. Then we have 
\begin{align}
\bbE \left[\|\bS\|^2\right]&= \bbE \left[\sum_{i=1}^n S_i^2 \right]\\
	&= \frac{1}{N}\sum_{\tilde{m}=1}^N\bbE \left[\sum_{i=1}^n G_i^2X_i^2 (\tilde{m},G^i)\right]\\
	&= \sum_{i=1}^n \bbE_{G} \left[G^2 \frac{1}{N}\sum_{\tilde{m}=1}^N \bbE \left[X_i^2 (\tilde{m},G^i)|G_i=G\right]\right]\\
	&= \sum_{i=1}^n \bbE_{G} \left[G^2 \varphi_i(G)\right]\\
	&= \bbE_{G} \left[G^2 \sum_{i=1}^n \varphi_i(G)\right]\\
	&= n\bbE_{G} \left[G^2 \varphi(G)\right]\\
	&\le n \Lambda.\label{Spowerconstrn}
\end{align}
Therefore, this choice of jammer satisfies the adversary power constraint. 
Given $\bY = \bg\circ\bx(M,\bg)+ \bg\circ \bx(\tilde{M},\bg)+\bV$, the decoder cannot determine the correct message between true message $M$ or the adversary message $\tilde{M}$ with high probability. Thus, the probability of error is bounded away from zero. By the above argument, if $ \mathbb{E} G^2 \varphi(G) \leq \Lambda$ for all $\varphi(g)$ where $\mathbb{E} \varphi(G)\le P$, then the capacity cannot be positive; the adversary can always symmetrize the channel, so the capacity is 0.  

On the other hand, consider the case where there exists some function $\varphi(g)$ where $\mathbb{E}G^2 \varphi(G)>\Lambda$ and $\mathbb{E} \varphi(G)\le P$.
Let $\psi_i(g)$ be given by
	\begin{align}
	\psi_i(g)= \argmin_{\psi(g): \bbE \psi(G)\le \Lambda} \bbE \left[C\left(\frac{G^2\varphi_i(G)}{\sigma^2+\psi(G)}\right)\right].
	\end{align}
Since the transmitted codes should work for arbitrary jamming signals, an outer bound may be obtained by assuming the adversary sends $S_i\sim \calN(0,\psi_i(G))$. By the assumption that $\mathbb{E} \psi_i(G)\le~\Lambda$, the jammer's expected power constraint is satisfied. Therefore, the rate is upper bounded by
\begin{align}
nR &\le \sum_{i=1}^n I(X_i;Y_i|G_i)\\
& = \sum_{i=1}^n I(X_i;G_i X_i+S_i+V_i|G_i)\\
& \le \sum_{i=1}^n \bbE_{G_i}\left[ C\left(\frac{G_i^2\varphi_i(G_i)}{\psi_i(G_i)+\sigma^2}\right)\right]\label{GFC1}\\
& = \sum_{i=1}^n \min_{\psi(g):\bbE\psi(g)\le \Lambda}\bbE_{G}\left[ C\left(\frac{G^2\varphi_i(G)}{\psi(G)+\sigma^2}\right)\right]\label{GFC2}\\
& \le n \min_{\psi(g):\bbE\psi(g)\le \Lambda}\bbE_{G} \left[C\left(\frac{G^2\frac{1}{n}\sum_{i=1}^n\varphi_i(G)}{\psi(G)+\sigma^2}\right)\right]\label{GFC3}\\
& \le n \underset{\substack{\varphi(g):\bbE \varphi(G)\leq P\\ \bbE G^2\varphi(G)\geq\Lambda}}{\max} \ \min_{\psi(g):\bbE\psi(g)\le \Lambda}\bbE_{G} \left[C\left(\frac{G^2\varphi(G)}{\psi(G)+\sigma^2}\right)\right]\label{GFC5}
\end{align}
where \eqref{GFC1} follows since the mutual information is less than the capacity of equivalent standard fading channel with noise variance $\psi_i(g_i)+\sigma^2$, and the gains being available at both encoder and decoder, \eqref{GFC2} follows by the definition of $\psi_i(g)$, \eqref{GFC3} follows by the concavity of $C(\cdot)$ with respect to $\varphi_i(g)$ and Jensen's inequality, and \eqref{GFC5} follows since we have established that $\varphi(g)=\frac{1}{n}\sum_{i=1}^n\varphi_i(g)$ satisfies $\bbE \varphi(G)\le P$ and $\bbE G^2\varphi(G)\ge \Lambda$.

Moreover, if the encoder knows the channel gains causally, and the adversary knows them non-causally, then the adversary is stronger than in the previous case, so exactly the same bound holds. If both encoder and adversary know the channel gains non-causally, then we instead assume 
\begin{align}
\varphi_i(g)=\frac{1}{N}\sum_{m=1}^N \bbE \left[ X_i^2(m,\bG)|G_i=g\right]
\end{align}
where $\bG = (G_1,\ldots,G_n)$ and $S_i = G_i X_i(\tilde{m},\bG)$, so we get the same upper bound. 

However, the case wherein the encoder knows the channel gains non-causally, and the adversary knows them causally is somewhat different. In this case, the encoder may send $ X_i^2(m,\bG)$ while the adversary does not have any access to $(G_{i+1}, \ldots, G_n)$ to construct $S_i = G_i X_i(\tilde{m},\bG)$. Thus, it cannot do better than sending Gaussian noise. In this case, we derive only a converse bound based on the adversary sending Gaussian noise. Hence, we obtain the following bound:
\begin{align}
R \le \underset{\substack{\varphi(g):\bbE \varphi(G)\leq P}}{\max} \ \min_{\psi(g):\bbE\psi(g)\le \Lambda}\bbE_{G} \left[C\left(\frac{G^2\varphi(G)}{\psi(G)+\sigma^2}\right)\right].
\end{align}
Note that here we are not making the assumption that $ \bbE G^2\varphi(G)\geq\Lambda$.

\subsection{Achievability Proof (Gains Available Non-causally at Adversary and Causally at Encoder)}\label{subsecVIII-B}

We first quantize $G$ in the following way. Fix $\nu>0$. Given the assumption that $G$ has finite variance, there exists a real-valued random variable $\tilde{G}$ with a finite support such that $\tilde{G}$ is a deterministic function of $G$ and $\bbE [(G-\tilde{G})^2|\tilde{G}=\tilde{g}]\le \nu$ for each $\tilde{g}$. We further assume that $\tilde{G}$ is the expected value of $G$ within each quantization set; that is, $\mathbb{E}[G|\tilde{G}]=\tilde{G}$.

Without loss of generality, assume $P=1$. Let $R$ be a rate and $\varphi(\tilde{g})$ be any function satisfying
\begin{align}
& \bbE \varphi(\tilde{G}) \le 1, \label{firstPowCond}\\
	&\Lambda<\bbE \tilde{G}^2\varphi(\tilde{G}), \label{eq:lambda_assumption4} \\ 
	&R<\underset{\psi(\tilde{g}):\bbE \psi(\tilde{G})\leq \Lambda}{\min}\bbE_{\tilde{G}} \left[C\left(\frac{\tilde{G}^2\varphi(\tilde{G})}{\psi(\tilde{G})+\sigma^2}\right)\right]. \label{eq:R_assumption4}
	\end{align}
We construct a $(2^{nR},n)$ code as follows:

\emph{Codebook generation:} 
Fix $\eps >\ \eps''> \eps ' >\lambda >0$. Generate $2^{nR}$ i.i.d. zero mean Gaussian
sequences $\bX(m)$ with variance $(1-\gamma )$ for each $m \in [2^{nR}]$. By Lemmas \ref{lem5} and Lemma \ref{lem5_0}, we may assume that the deterministic codebook satisfies \eqref{eq:two_codebooks_c0}--\eqref{eq:21Huz_c4}.

\emph{Encoding:} Given message $m$ and gain sequence $\bg$, the transmitter computes $\tilde{g}$ from the quantization function, and then sends $\sqrt{\varphi(\tilde{\bg})}\circ\bx(m)$ (at time $i$ signal $\sqrt{\varphi(\tilde{g_i})}x_i(m)$ is sent) if $\|\sqrt{\varphi(\tilde{\bg})}\circ\bx(m)\|^2\le n$; otherwise, it sends zero. Note that here we assume that the encoder knows the channel gains causally. 

\emph{Decoding:} Given $\by$ and $\bg$, let $\nu< \eps$ and $\mathscr{S}$ be the set of messages $\hat{m}$ such that $(\bx(\hat{m}),\tilde{\bg}, \by)\in \calT_\eps^{(n)}(X',\tilde{G},Y)$ where $\tilde{G}$ is the quantized random variable from $G$ and $(X',Y)$ are some random variables that are conditionally Gaussian given $\tilde{G}=\tilde{g}$ with zero mean and covariance
\begin{align}\label{sdefinition2}
\cov\left(X',Y\Big|\tilde{G}=\tilde{g}\right)=\left[\begin{array}{cccc}1&\tilde{g}\sqrt{\varphi(\tilde{g})} \\\tilde{g}\sqrt{\varphi(\tilde{g})}&a_{\tilde{g}}\end{array}\right]
\end{align}
where $a_{\tilde{g}}\ge \tilde{g}^2 \varphi(\tilde{g})+\sigma^2$. Note that the following can be shown from \eqref{sdefinition2}.
\begin{gather}
X' \text{ is independent of } \tilde{G}, \label{s15}\\
\bbE X'^2 = 1,\label{s25}\\
Y-\tilde{G}\sqrt{\varphi(\tilde{G})}X' \text{ is independent of } X' \text{ given } \tilde{G},\label{s35}\\
\var\left(Y-\tilde{G}\sqrt{\varphi(\tilde{G})}X'\bigg|\tilde{G}\right) \ge \sigma^2. \label{s45}
\end{gather}
Now, we define the decoding function as
\begin{align}
	\Theta(\mathbf{y},\tilde{\bg}) &= \argmin_{\hat{m} \in  \mathscr{S}}
 \left\| \by - \tilde{\bg} \circ \sqrt{\varphi(\tilde{\bg})} \circ  \bx(\hat{m}) \right\|^2.
\end{align}

\emph{Analysis of the probability of error:} Assume the legitimate transmitter sends message $M$. Then, we can upper bound the probability of error by the summation of the following error probabilities:
\begin{align}
P_0&=\bbP\left\{M \notin\mathscr{S} \right\},\label{firstPe}\\
P_1&=\bbP\bigg\{\left\| \bY -\tilde{ \bG}\circ \sqrt{\varphi(\tilde{\bG})}\circ \bx(\hat{m}) \right\|^2
\leq \left\| \bs+\bV \right\|^2 \text{ for some } \hat{m}\in \mathscr{S}\setminus \{M\} \bigg\}.\label{secondPe}
\end{align}

We can prove with high probability that
\begin{align}
\frac{1}{n}\left\|\bx\circ \sqrt{\varphi(\tilde{\bG})}\circ(\bG-\tilde{\bG})\right\|^2 &=  \frac{1}{n} \sum_{i=1}^n \left(x_i \sqrt{\varphi(\tilde{G}_i)} \left(G_i - \tilde{G}_i\right)\right)^2\\
& \le \frac{1}{n} \sum_{i=1}^n x_i^2 \bbE_{\tilde{G}_i} \left[\bbE_{G_i} \left[\varphi(\tilde{G}_i)\left(G_i - \tilde{G}_i\right)^2\Big| \tilde{G}_i \right]\right] + \nu\label{bynu111}\\
& \le \frac{1}{n} \sum_{i=1}^n x_i^2 \bbE_{\tilde{G}_i} \left[\varphi(\tilde{G}_i)\right] \nu + \nu\label{bynu211}\\
& \le 2\nu\label{bynu123}
\end{align}
where \eqref{bynu111} follows from the law of large numbers for non-identical independent random variables $x_i^2\varphi(\tilde{G}_i)\left(G_i-\tilde{G}_i\right)^2$, \eqref{bynu211} follows from the assumption $\bbE \left[\left(G_i - \tilde{G}_i\right)^2\Big| \tilde{G}_i=\tilde{g}_i\right] \leq \nu$, and \eqref{bynu123} follows from the assumption $\frac{1}{n} \|\bx\|^2\leq 1$ and $\bbE \varphi(\tilde{G}) \leq 1$.

Consider any jammer sequence $\mathbf{s}$. We may assume sequence $\bG$ is typical since it is drawn i.i.d. from the distribution $f_G(g)$. Similarly, $\tilde{\bG}$ is also typical because it is from the corresponding discrete distribution $P_{\tilde{G}}(\tilde{g})$. Thus, $(\bs,\tilde{\bG})$ is also typical with respect to some distribution $P_{\tilde{G}}(\tilde{g})f_{S|\tilde{G}}(s|\tilde{g})$ where $f_{S|\tilde{G}}(s|\tilde{g})$ is conditionally Gaussian. Note that we can make no assumptions about the conditional variances defining $f_{S|\tilde{G}}$, because the adversary is assumed to know $G$ in its choice of $s$. By \eqref{eq:two_codebooks_c0}, with high probability $(\mathbf{x}(M),\mathbf{s},\tilde{\bG})\in \mathcal{T}_{\eps'}^{(n)}(X,S,\tilde{G})$ where $X$ is independent of $(S,\tilde{G})$, and $\bbE X^2=1 ,\allowbreak \bbE S^2 \le \Lambda$.
Thus, by the conditional typicality lemma, with high probability $(\bx,\bs,\tilde{\bG},\bV)\in \calT_{\eps''}^{(n)}(X,S,\tilde{G},V)$ where $X,S,\tilde{G}$ are independent of $V$, and $\bbE V^2=\sigma^2$. Hence, using \ref{bynu123}, we have $\left(\bx,\bs,\tilde{\bG},\bV +\bx \circ (\bG - \tilde{\bG}) \circ \sqrt{\varphi(\tilde{\bG})}\right)\in \T(X,S,\tilde{G},V)$ where $\nu$ is sufficiently small compared to $\eps$. Also, since $\bY-\bx \circ \tilde{\bG} \circ \sqrt{\varphi(\tilde{\bG})}-\bs -\bV = \bx \circ (\bG - \tilde{\bG} )\circ \sqrt{\varphi(\tilde{\bG})}$, by \eqref{bynu123} we may roughly assume $\bY=\bx \circ \tilde{\bG} \circ \sqrt{\varphi(\tilde{\bG})}+\bs +\bV$ and obtain $\left(\bx,\bs,\tilde{\bG},\bY\right)\in \T(X,S,\tilde{G},Y)$. Moreover, to completely show that with high probability $M\in\mathscr{S}$, we also need to compute the covariance matrix of $(X,Y)$ given $\tilde{G}=\tilde{g}$, where $Y=\tilde{G}\sqrt{\varphi(\tilde{G})} X+S+V$, and show that it is in the form of \eqref{sdefinition2}.
First, $\bbE \left(X^2|\tilde{G}=\tilde{g}\right) = \bbE X^2 = 1$ since $X$ is independent of $\tilde{G}$, 
\begin{align}
\bbE \left(\!X\!\left(\!\tilde{G}\sqrt{\varphi(\tilde{G})}X\!+\!S\!+\!V\right)\Big|\tilde{G}\!=\!\tilde{g}\right) \!
& =\tilde{g}\sqrt{\varphi(\tilde{g})}\bbE X^2\! +\!  \bbE \left(XS|\tilde{G}\! =\! \tilde{g}\right)\! +\! \bbE \left(XV|\tilde{G}\! =\! \tilde{g}\right)\! \\
& =\tilde{g}\sqrt{\varphi(\tilde{g})}\label{cov_22}
\end{align}
where $ \bbE \left(XS\Big|\tilde{G}=\tilde{g}\right)=0$ follows from the weak union rule since $X$ is independent of $(S,G)$. 
Furthermore,

\begin{align}
\bbE &\left(\left(\tilde{G}\sqrt{\varphi(\tilde{G})}X+S+V\right)^2\bigg|\tilde{G}=\tilde{g}\right)
 = \bbE \left(\tilde{G}^2{\varphi(\tilde{G})}X^2\bigg|\tilde{G}=\tilde{g}\right)+\bbE \left(S^2\Big|\tilde{G}=\tilde{g}\right)\nonumber\\
 &+\bbE \left(V^2\Big|\tilde{G}=\tilde{g}\right) +\!2\bbE \left(\tilde{G}\sqrt{\varphi(\tilde{G})}XS\bigg|\tilde{G}\!=\!\tilde{g}\right)\!+\!2\bbE \left(\tilde{G}X\sqrt{\varphi(\tilde{G})}V\bigg|\tilde{G}\!=\!\tilde{g}\right)\!+\! 2\bbE \left(SV\Big|\tilde{G}\!=\!\tilde{g}\right)\\
&= \tilde{g}^2{\varphi(\tilde{g})}\!+\!\bbE \left(S^2\Big|\tilde{G}\!=\!\tilde{g}\right)\!+\!\sigma^2\!+\!2\tilde{g}\sqrt{\varphi(\tilde{g})}\bbE \left(XS\Big|\tilde{G}\!=\!\tilde{g}\right)\!+\!2\tilde{g}\sqrt{\varphi(\tilde{g})}\bbE \left(XV\Big|\tilde{G}\!=\!\tilde{g}\right)\!\nonumber\\
&\quad{}+\! 2\bbE \left(SV\Big|\tilde{G}\!=\!\tilde{g}\right)\\
&= \tilde{g}^2{\varphi(\tilde{g})}+\bbE \left(S^2\Big|\tilde{G}=\tilde{g}\right)+\sigma^2\label{weekunion3} \\
&\geq \tilde{g}^2{\varphi(\tilde{g})}+\sigma^2\label{cov_23}
\end{align}
where \eqref{weekunion3} follows from the weak union rule for $X$ independent of $(S,\tilde{G})$ and $V$ independent of $(S,\tilde{G})$. Therefore, the conditional covariance matrix of $(X,Y)$ can be obtain from $\bbE X^2 = 1$, \eqref{cov_22} and \eqref{cov_23}, and is the same as \eqref{sdefinition2}. Now, since $(\bx(\hat{M}),\tilde{\bg}, \by)\in \calT_\eps^{(n)}(X,\tilde{G},Y)$ and the conditional covariance matrix of $(X(M),Y)$ satisfies \eqref{sdefinition2}, with high probability $M\in \mathscr{S}$, and $P_0$ vanishes as $n\to\infty$.

Using \eqref{bynu123} and triangle inequality, we may upper bound $P_1$ by the following:
\begin{multline}
P_1\leq \\
\mathbb{P}\left\{ \!\left\|\bx(m)\!\circ\! \tilde{\bG}\sqrt{\varphi(\tilde{\bG})}\!+\!\bs\!+\!\bV\!-\!\bx(\hat{m})\!\circ\! \tilde{\bG}\sqrt{\varphi(\tilde{\bG})}\right\|^2\!\!\!\le\! \|\bs\!+\!\bV\|^2 \!+\!2n \nu
\text{ for some }\! \hat{m}\!\in\!\mathscr{S}\setminus\! \{m\}\!\right\}.
\end{multline}
Define the shorthand $\vec{X}=(XX'S\tilde{G}V)$. Let $\mathcal{V}$ denote a finite $\eps$-dense subset in the set of all distributions of random vectors $\vec{X}$ that are determined by $P_{\tilde{G}}(\tilde{g})$ and a random vector $(XX'SV)$ distributed conditionally zero mean Gaussian given $\tilde{G}$ with bounded covariances at most $(1,1,\Lambda,\sigma^2)$. Note that because the distribution of $P_{\tilde{G}}(\tilde{g})$ is completely known, the overall distribution of $\vec{X}$ can be determined by the conditional covariance matrix of $(XX'SV)$ given $\tilde{G}=\tilde{g}$ for each of the finitely many $\tilde{g}$ realizations, so $\mathcal{V}$ only needs to cover a compact set. Now, we may upper bound $P_1$ by
\begin{equation}
\sum_{\vec{X}\in\mathcal{V}} \frac{1}{N} \sum_{m=1}^N \bbE_{\tilde{G}}\left[e_{\vec{X}}(m,\mathbf{s},\tilde{\bG})\right]
\end{equation}
where 
\begin{multline}
e_{\vec{X}}\left(m,\mathbf{s},\tilde{\bg}\right)
= \mathbb{P}\bigg\{ \left(\bx(m),\bx(\hat{m}),\bs,\tilde{\bg},\bV\right)\in\calT_\eps^{(n)}\left(\vec{X}\right),\\
\left\|\tilde{\bg}\!\circ\! \sqrt{\varphi(\tilde{\bg})}\!\circ\! \bx(m)\!+\!\bs\!+\!\bV\!-\!\tilde{\bg}\!\circ\! \sqrt{\varphi(\tilde{\bg})}\!\circ\! \bx(\hat{m})\right\|^2\!\le \!\|\bs\!+\!\bV\|^2 \!+\!2n \nu
\text{ for some }\ \hat{m}\in\mathscr{S}\setminus \{m\}\bigg\}.\label{nu_eq5}
\end{multline}

We will show that $\frac{1}{N} \sum_{m=1}^N e_{\vec{X}}(m,\mathbf{s},\tilde{\bg})\to 0$ for all vectors $\tilde{\bg}$ and all vectors $(XX'SV)$ which are Gaussian given $\tilde{G}$ (whether or not they are in $\calV$). 
Let $Z = \tilde{G}\sqrt{\varphi(\tilde{G})}X+S+V-\tilde{G}\sqrt{\varphi(\tilde{G})}X'$.
We may restrict ourselves to $\vec{X}$ where
\begin{gather}
I(X;X'S\tilde{G})< |R-I(X';S\tilde{G})|^++\delta(\eps),\label{eq:R1R25}\\
\tilde{G}\sim P_{\tilde{G}}(\tilde{g}),\label{signalpower5}\\
(X,X',S,V)\text{ are zero mean Gaussian given }\tilde{G},\label{defofS5}\\
X,(S,\tilde{G}),V\text{ are mutually independent},\label{XSV_indep5}\\
X', \tilde{G}\text{ are independent}, \label{X'indG5}\\
\bbE X^2=\bbE X'^2=1, \bbE S^2 \le \Lambda, \bbE V^2=\sigma^2,\label{codgenr5}\\
X',Z \text{ are independent given } \tilde{G},\label{XZ1_indep5}\\
\bbE \left[Z^2\Big|\tilde{G}\right]\ge \sigma^2,\label{Zles5}\\
\var( Z)\le \sigma^2+\Lambda+2\nu,\label{Zless5}.
\end{gather}
Note that using \eqref{eq:21Huz_c4}, we only need to consider the distributions that satisfies \eqref{eq:R1R25}. in addition, \eqref{signalpower5}--\eqref{defofS5} are obtained by the definition of $\mathscr{S}$, \eqref{XSV_indep5} holds since the codebook $X$, Gaussian noise $V$ and fading gains $\tilde{G}$ are generated independently, and the adversary signal $S$ may depend on $\tilde{G}$ but not the others, \eqref{X'indG5} follows from \eqref{s15}, \eqref{codgenr5} follows from the power constraints of the codebook, the adversary and the distribution of noise, \eqref{XZ1_indep5}-\eqref{Zles5} follows from \eqref{s35}-\eqref{s45}, and \eqref{Zless5} follows from \eqref{nu_eq5}. Let $\psi(\tilde{g})=\bbE\left[Z^2\Big|\tilde{G}=\tilde{g}\right]-\sigma^2$. Therefore, using \eqref{Zles5} we have $\psi(\tilde{g})\ge 0$, and by \eqref{Zless5} we get $\bbE \psi(\tilde{G})=\var(Z)-\sigma^2\le \Lambda+2\nu$. 

Observe that if $I(XV;X'S|\tilde{G})=0$, then we would have
\begin{align}
0&= \bbE \left[X'Z|\tilde{G}\right]\label{XZindependent}\\
 &= \bbE \left[X'\left(\tilde{G}\sqrt{\varphi(\tilde{G})}X+S+V-\tilde{G}\sqrt{\varphi(\tilde{G})}X'\right)\bigg|\tilde{G}\right]\\
&= \bbE \left[X'S\Big|\tilde{G}\right] -\bbE \left[\tilde{G}\sqrt{\varphi(\tilde{G})}X'^2\bigg|\tilde{G}\right]\label{mutual1}\\
&= \bbE \left[X'S\Big|\tilde{G}\right] -\tilde{G}\sqrt{\varphi(\tilde{G})}\label{XindepntG}
\end{align}
where \eqref{XZindependent} follows from \eqref{XZ1_indep5}, \eqref{mutual1} follows from the assumption $I(XV;X'S|\tilde{G})=0$ in which $X'$ is independent of $(X,V)$, and \eqref{XindepntG} holds since $X'$ is independent of $\tilde{G}$. Therefore, $\bbE \left[X'S\Big|\tilde{G}\right] = \tilde{G}\sqrt{\varphi(\tilde{G})}$ and the covariance matrix of $S,X'$ given $\tilde{G}$ is equal to
\begin{align}
	\cov\left(S,X'\Big|\tilde{G}\right)=\begin{bmatrix}
	\bbE \left[S^2\Big|\tilde{G}\right]& \tilde{G}\sqrt{\varphi(\tilde{G})} \\
	\tilde{G}\sqrt{\varphi(\tilde{G})}& 1
	\end{bmatrix}.
\end{align}
The determinant of $\cov\left(S,X'\Big|\tilde{G}\right)$ is $\bbE\left[S^2\Big|\tilde{G}\right] -\tilde{G}^2\varphi(\tilde{G})$ that should be non-negative since the covariance matrix must be positive semi-definite. Thus, its expectation is also non-negative:
\begin{align}\label{psd}
	0 & \le \bbE S^2 - \bbE \tilde{G}^2\varphi(\tilde{G}).
\end{align}
However, since $\mathbb{E}S^2 \leq \Lambda$, \eqref{psd} contradicts the initial assumption on $\varphi$ in \eqref{eq:lambda_assumption4}. Thus, there exists $\eta>0$ such that
\be\label{eq:eta_bound5}
\eta\le I(XV;X'S|\tilde{G})=I(XV;X'|S\tilde{G})
\ee
where we have used the fact that $I(XV;S)=0$.

Probability $e_{\vec X}$ may be upper bounded by
\begin{align}
e_{\vec X}(m,\bs,\tilde{\bg})&\leq\sum_{\hat{m}:(\bx(m),\bx(\hat{m}),\bs,\tilde{\bg})\in\T(X,X',S,\tilde{G})}\hspace{-2em}\bbP\left\{\!(\bx(m),\bx(\hat{m}),\bs,\tilde{\bg},\bV)\!\in\! \T (X,X',S,\tilde{G},V)\!\right\}
\\&\leq \exp\big\{n\big[|R\!-\! I(X';XS\tilde{G})|^+\!\!-I(V;X'|XS\tilde{G})\!+\!\delta(\eps)\big]\label{e(s,i)5}
\end{align}
where \eqref{e(s,i)5} follows from \eqref{eq:22Huz_c5} and the joint typicality lemma.

We consider the following two cases.

Case (a): $R<I(X';S\tilde{G})$. Applying this condition to \eqref{eq:R1R25}, we get 
\begin{align}
\delta(\eps)& >I(X;X'S\tilde{G}) \\
& = I(X;X'|S\tilde{G}).\label{IXX'2}
\end{align}
Since $I(X';S\tilde{G})\le I(X';XS\tilde{G})$ then $R-I(X';XS\tilde{G})< 0$. Considering \eqref{e(s,i)5}, for any $m, \bs,\tilde{\bg}$ we have
\begin{align}
e_{\vec X}(m,\bs,\tilde{\bg})&\le \exp\left\{-n\left( I(V;X'|XS\tilde{G})-\delta(\eps)\right)\right\}\label{eq:two_codebook_ebound5}\\
&= \exp\{- n(I(XV;X'|S\tilde{G})-I(X;X'|S\tilde{G})-\delta(\eps))\}\\
&\le \exp\{-n(\eta-2\delta(\eps))\}\label{casearesult5}
\end{align}
where \eqref{casearesult5} follows from \eqref{eq:eta_bound5} and \eqref{IXX'2}. Therefore, $e_{\vec X}(m,\bs,\tilde{\bg})$ vanishes exponentially fast if $\delta(\eps)$ is sufficiently small.

Case (b): $R\geq I(X';S\tilde{G})$. Then we may apply this condition to \eqref{eq:R1R25} as
\begin{align}
	R &> I(X;X'S\tilde{G})+I(X';S\tilde{G})-\delta(\eps)\\
	& \ge I(X;X'|S\tilde{G})+I(X';S\tilde{G})-\delta(\eps)\\
	& =I(X';XS\tilde{G})-\delta(\eps).
\end{align}
Since $R-I(X';XS\tilde{G})+\delta(\eps)>0$, we may upper bound \eqref{e(s,i)5} by 
\begin{align}
	e_{\vec X}(m,\bs,\tilde{\bg})
	& \leq \exp \left(n\left[R\!-\! I(X';XS\tilde{G})\!-\! I(V;X'|XS\tilde{G})\!+\!2\delta(\eps)\right]\right)\label{formula1}\\
& = \!\exp (n[R-I(X';XS\tilde{G}V)+2\delta(\eps)])\\
& \le \!\exp (n[R-I(X';XSV|\tilde{G})+2\delta(\eps)])\label{result335}
\end{align}
where by \eqref{X'indG5}, we have $I(X';\tilde{G})=0$. In the following, we find a lower bound for the mutual information in \eqref{result335}.
\begin{align}
I\left(X';XSV\Big|\tilde{G}\right)
& \geq I\left(X';\tilde{G}\sqrt{\varphi(\tilde{G})}X\!+\!S\!\!+\!\!V\Big|\tilde{G}\right)\label{dataproces2}\\
& = I\left(X';Z+\tilde{G}\sqrt{\varphi(\tilde{G})}X'\Big|\tilde{G}\right)\\
& = h\left(Z+\tilde{G}\sqrt{\varphi(\tilde{G})}X'\Big|\tilde{G}\right)-h\left(Z+\tilde{G}\sqrt{\varphi(\tilde{G})}X'\Big|\tilde{G},X'\right)\\
& = \bbE_{\tilde{G}} \left[\frac{1}{2} \log 2\pi e\left(\tilde{G}^2{\varphi(\tilde{G})}+\bbE \left[Z^2\Big|\tilde{G}\right]\right)-\frac{1}{2} \log 2\pi e \bbE \left[Z^2\Big|\tilde{G}\right] \right] \\
& = \bbE_{\tilde{G}} \left[C\left(\frac{\tilde{G}^2\varphi(\tilde{G})}{\bbE \left[Z^2\Big|\tilde{G}\right]} \right) \right] \label{GaussCap5}\\
& = \bbE_{\tilde{G}} \left[C\left(\frac{\tilde{G}^2\varphi(\tilde{G})}{\psi(\tilde{G})+\sigma^2+2\nu}\right)\right]\label{psiDef}
\end{align}
where \eqref{dataproces2} follows from data processing inequality, \eqref{GaussCap5} follows from standard argument for the capacity of Gaussian channel, and \eqref{psiDef} follows from the definition of $\psi$. Therefore, by the assumptions about $R$ and $\Lambda$ in \eqref{eq:lambda_assumption4}--\eqref{eq:R_assumption4}, $R<I(X';XSV|\tilde{G})$, so by \eqref{result335} $e_{\vec X}(m,\bs,\tilde{\bg})$ is exponentially vanishing if $\delta(\eps)$ and $\nu$ are sufficiently small.

It is worth mentioning that this achievability proof also works for the case where both the adversary and encoder know the channel gains causally, or both know the gains non-causally. Since in all three cases the knowledge of the encoder is not more than the knowledge of the adversary, the jammer is able to impersonate the legitimate transmitter, and thereby symmetrize the channel, depending on the power allocation.

\subsection{Achievability Proof (Gains Available Causally at Adversary and Non-causally at Encoder)}\label{subsecVIII-C}

In this case, both the encoder and the decoder know the channel gains non-causally meaning that they know the whole $\bg$ string including $(g_1,g_2,\cdots,g_n)$. However, the adversary only knows the gains causally, so at time $i$ it only has access to $(g_1,g_2,\cdots,g_i)$. Therefore, both the encoder and the decoder have some extra common information $(g_{i+1},g_{i+2},\cdots,g_n)$ that the adversary does not know. In particular, the encoder and the decoder immediately know $g_n$ which the adversary knows only at time $n$. Hence, we can leverage this common knowledge between the encoder and the decoder as common randomness that is unknown to the jammer. Moreover, by the assumption that $G$ is a continuous random variable with positive variance, in fact just $G_n$ has infinite entropy, and thus can be considered a source of an infinite number of bits of common randomness. Therefore, we proceed to provide an achievability proof where the encoder and decoder are assumed to share an infinite source of common randomness. However, note that implementing this approach would require measuring $G_n$ to an arbitrarily level of precision, which is not practical. Even so, the random code reduction technique of, for example, \cite[Lemma 12.8]{CsiszarKorner}, can be used to show that only $O(\log n)$ bits of common randomness need to be extracted from $G_n$ (or perhaps $G_{n-k},\ldots, G_n$ for some $k$) in order to achieve the same rate.

A large amount of the achievability proof is identical to the Sec. \ref{subsecVIII-B}. The main difference is that the codebook is based on common randomness between encoder and decoder, so we denote the codebook as random variables $\mathbf{X}(m)$ which are independent from the jammer signal. As a consequence, the symmetrizability condition $\Lambda<\bbE \tilde{G}^2\varphi(\tilde{G})$ is not needed, and the proof is somewhat simpler. In particular, we fix a $\varphi(\cdot)$ satisfying \eqref{firstPowCond}, and a rate satisfying \eqref{eq:R_assumption4}, and we prove achievability as follows.
 
\emph{Codebook generation:} 
Let $\bX(m)$ be a Gaussian codebook with variance $1\!-\!\gamma$ satisfying \eqref{eq:two_codebooks_c0}. This random codebook is generated from the infinite source of common randomness, so it is unknown to the adversary.

\emph{Encoding:} Given message $m$ and gain sequence $\bg$, the transmitter first computes $\tilde{g}$ from the quantization function, and then sends $\sqrt{\varphi(\tilde{\bg})}\circ\bX(m)$ (at time $i$ signal $\sqrt{\varphi(\tilde{g_i})}X_i(m)$ is sent) if $\|\sqrt{\varphi(\tilde{\bg})}\circ\bX(m)\|^2 \le n$; otherwise, it sends zero.  

\emph{Decoding:} Given $\by$ and $\bg$, let $\nu< \eps$ and let $\mathscr{S}$ be the set of messages $\hat{m}$ such that $(\bX(\hat{m}),\tilde{\bg}, \by)\in \calT_\eps^{(n)}(X',\tilde{G},Y)$ where $\tilde{G}$ is the quantized random variable from $G$ and $(X',Y)$ are conditionally Gaussian given $\tilde{G}=\tilde{g}$ with zero mean and covariance matrix  $\Sigma_{\tilde{g}}$ as follows: 
\begin{align}\label{sdefinition}
\Sigma_{\tilde{g}}=\cov\left(X',Y\Big|\tilde{G}=\tilde{g}\right)=\left[\begin{array}{cccc}1&\tilde{g}\sqrt{\varphi(\tilde{g})} \\\tilde{g}\sqrt{\varphi(\tilde{g})}&a_{\tilde{g}}\end{array}\right]
\end{align}
where $a_{\tilde{g}}\ge {\tilde{g}}^2 \varphi({\tilde{g}})+\sigma^2$. Note that the following can be shown from \eqref{sdefinition}:
\begin{gather}
X' \text{ is independent of } {\tilde{G}}, \label{s1}\\
\bbE X'^2 = 1,\label{s2}\\
Y-{\tilde{G}}\sqrt{\varphi({\tilde{G}})}X' \text{ is independent of } X' \text{ given } {\tilde{G}},\label{s3}\\
\var\left(Y-{\tilde{G}}\sqrt{\varphi({\tilde{G}})}X'\bigg|{\tilde{G}}\right) \ge \sigma^2.\label{s4}
\end{gather}
Now, we define the decoding function as
\begin{align}
\Theta(\mathbf{y},\tilde{\bg}) &= \argmin_{\hat{m} \in  \mathscr{S}}
\left\| \by - \tilde{\bg} \circ \sqrt{\varphi(\tilde{\bg})} \circ  \bX(\hat{m}) \right\|^2.
\end{align}

\emph{Analysis of the probability of error:} Assume the legitimate transmitter sends message $M$. Then, we can upper bound the probability of error by the summation of the following error probabilities:
\begin{align}
P_0&=\bbP\left\{M \notin\mathscr{S} \right\},\label{firstPe}\\
P_1&=\bbP\bigg\{\left\| \bY - \tilde{\bG}\circ \sqrt{\varphi(\tilde{\bG})}\circ \bX(\hat{m}) \right\|^2
\leq \left\| \bs+\bV \right\|^2 \text{ for some } \hat{m}\in \mathscr{S}\setminus \{M\} \bigg\}.
\label{secondPe}
\end{align}

Using the same argument in Sec. \ref{subsecVIII-B}, we can prove with high probability that $P_0$ tends to zero as $n\to \infty$. Using \eqref{bynu123} and triangle inequality, we may upper bound $P_1$ by the following:
\begin{multline}
P_1\leq \\
\mathbb{P}\left\{ \!\left\|\bX(m)\!\circ\! \tilde{\bG}\sqrt{\varphi(\tilde{\bG})}\!+\!\bs\!+\!\bV\!-\!\bX(\hat{m})\!\circ\! \tilde{\bG}\sqrt{\varphi(\tilde{\bG})}\right\|^2\!\!\!\le\! \|\bs\!+\!\bV\|^2 \!+\!2n \nu
\text{ for some }\! \hat{m}\!\in\!\mathscr{S}\setminus\! \{m\}\!\!\right\}.
\end{multline}
Defining $\vec{X}=(XX'S\tilde{G}V)$ and $\mathcal{V}$ the same as Sec. \ref{subsecVIII-B}, we may now upper bound $P_1$ by
\begin{equation}
\sum_{\vec{X}\in\mathcal{V}} \frac{1}{N} \sum_{m=1}^N \bbE_{{\tilde{G}}}\left[e_{\vec{X}}(m,\mathbf{s},\tilde{\bG})\right]
\end{equation}
where 
\begin{multline}\label{eq:eX1_def}
e_{\vec{X}}(m,\mathbf{s},\tilde{\bg})
=\!\bigg\{\! (\bX(m),\bX(\hat{m}),\bs,\tilde{\bg},\bV)\!\in\!\calT_\eps^{(n)}(\vec{X}),\\
\quad{} \left\|\tilde{\bg}\!\circ\! \sqrt{\varphi(\tilde{\bg})}\!\circ\! \bX(m)\!+\!\bs\!+\!\bV\!-\!\tilde{\bg}\!\circ\! \sqrt{\varphi(\tilde{\bg})}\!\circ\!  \bX(\hat{m})\right\|^2\!\le\! \|\bs+\!\bV\|^2 + 2n \nu\text{ for some } \hat{m}\!\in\! \mathscr{S}\setminus \{M\}\!\bigg\},
\end{multline}
and $\vec{X}$ satisfies the same properties as in \eqref{eq:R1R25}--\eqref{Zless5}.
Now, it suffices to show that $\frac{1}{N} \sum_{m=1}^N e_{\vec{X}}(m,\mathbf{s},\tilde{\bg})$ vanishes for all typical vectors $\bg$ and all vectors $(XX'SV)$ which are Gaussian given ${\tilde{G}}$ (whether or not they are in $\calV$).

Using the joint typicality lemma in \cite[Remark 2.2]{ElGamal} we may upper bound $	e_{\vec X}$ in \eqref{nu_eq5} (with the codewords $\bX(m)$ and $\bX(\hat{m})$) as follows:
\begin{align}
	e_{\vec X}(m,\bs,\tilde{\bg})
	&\leq\sum_{\hat{m}\in \mathscr{S}\setminus \{m\}} \bbP\left\{\!(\bX(m),\bX(\hat{m}),\bs,\tilde{\bg},\bV)\!\in\! \T (X,X',S,{\tilde{G}},V)\!\right\}
	\\& \le \exp \{n(R-I(X';XSV{\tilde{G}})+\eps)\}\label{eq:174}\\
	& = \!\exp \{n(R-I(X';XSV|{\tilde{G}})+\eps)\}\label{result333}
\end{align}
where in \eqref{eq:174} we have used the fact that $\bX(\hat{m})$ is independent of $(\bX(m),\bs,\tilde{\bg},\bV)$, and \eqref{result333} follows from \eqref{s1}.
We now lower bound the mutual information in \eqref{result333} by the following.
\begin{align}
I(X';XSV|{\tilde{G}})
 & \geq I\left(X';{\tilde{G}}\sqrt{\varphi({\tilde{G}})}X+S+V\Big|{\tilde{G}}\right)\label{dataProcSS}\\
& = I\left(X';Z+{\tilde{G}}\sqrt{\varphi({\tilde{G}})}X'\Big|{\tilde{G}}\right)\\
& = h\left(Z+{\tilde{G}}\sqrt{\varphi({\tilde{G}})}X'\Big|{\tilde{G}}\right)-h\left(Z+{\tilde{G}}\sqrt{\varphi({\tilde{G}})}X'\Big|{\tilde{G}},X'\right)\\
& = \bbE_{\tilde{G}} \bigg[\frac{1}{2} \log 2\pi e\left({\tilde{G}}^2\varphi({\tilde{G}})\bbE \left[X'^2\Big|{\tilde{G}}\right]+\bbE \left[Z^2\Big|{\tilde{G}}\right]\right)-\frac{1}{2} \log 2\pi e \bbE \left[Z^2\Big|{\tilde{G}}\right] \bigg] \\
& = \bbE_{\tilde{G}} \left[C\left(\frac{{\tilde{G}}^2\varphi({\tilde{G}})}{\bbE \left[Z^2\Big|{\tilde{G}}\right]} \right) \right] \label{GaussCap}\\
& = \bbE_{\tilde{G}} \left[C\left(\frac{{\tilde{G}}^2\varphi({\tilde{G}})}{\psi({\tilde{G}})+\sigma^2+2\nu}\right)\right]\label{psiDef6}
\end{align}
where \eqref{dataProcSS} follows from data processing inequality, \eqref{GaussCap} follows from standard argument for the capacity of Gaussian channel, and \eqref{psiDef6} follows from the definition of $\psi$.  Therefore, by the assumptions about $R$ and $\Lambda$ in \eqref{firstPowCond} and \eqref{eq:R_assumption4}, $R<I(X';XSV|{\tilde{G}})$, so by \eqref{result333} $e_{\vec X}(m,\bs,\tilde{\bg})$ is exponentially vanishing if $\delta(\eps)$ and $\nu$ are sufficiently small.

\section{Conclusion}\label{9}

This paper studied two phenomena together which are usually studied separately: namely, active adversaries and fading. We derived the capacity of Gaussian arbitrarily-varying fading channels where the adversary knows the transmitter's code but not the exact transmitted signal. The adversary affects the capacity by increasing the noise variance by an amount related the adversary's power. The capacity also depends on whether the transmitter and/or the adversary know the fading gains or not. The transmitter uses its knowledge to maximize the channel capacity while the adversary tries to minimize the capacity by its knowledge of the channel gains. Furthermore, if the adversary's knowledge is at least that of the transmitter's knowledge, then the adversary is able to make the capacity zero with enough power. In this paper, we have focused on the scenario where fading applies to the transmitter-to-receiver path, but not the adversary-to-receiver path. Future work could including considering fading along both paths. Such an alternatively model would present somewhat different challenges. Alternative directions include considering fading and adversaries in network settings, or an adversary with some direct control of the fading gains.

\section*{Acknowledgment}
This material is based upon work supported by the National Science Foundation under Grant No. CCF-1453718.

\appendix

\section{Proof of Lemma \ref{lem5}}\label{Apn1}

In order to prove \eqref{eq:two_codebooks_c0}, we use our proof in \cite[Lemma 6]{Fatemeh2017} for one codebook. Moreover, to obtain \eqref{eq:22Huz_c5}--\eqref{eq:21Huz_c4}, we apply the corresponding proof of the equations in \cite[Lemma 1]{Hughes} for Gaussian distributions. Note that \cite{Hughes} focuses on discrete alphabets, but the same proofs can be extended to Gaussian distributions by quantization of the set of continuous random variables in the following way.

Let $\bX_{i}$ be Gaussian i.i.d. $n$-length random vectors (codebook) independent from each other with $\var(X) = 1$. First let $\bg\in \bbR^n$ be a typical realization of $n$ i.i.d. continuous random variable $G$ with probability density function $f_G(g)$. Next, we quantize the set of all $\bg \in \bbR^n$, into a $\nu$-dense subset $\calG^n$. For a fixed $\bg \in \calG^n$, fix $\bx\in \calT_\eps^{(n)}(X), \bs \in \scU^n$ and a covariance matrix $\cov(X,X',S|G=g)\in\calV^{3\times 3}$ such that $\scU^n$ is a $\nu$-dense subset of $\bbR^n$ for $\bs$ such that $||\bs||^2\le n\Lambda$, and $\calV^{3\times 3}$ is a $\nu$-dense subset of $\bbR^{3\times 3}$ for positive definite covariance matrices with diagonals at most $(1,1,\Lambda)$.

Using the similar proof in \cite[Lemma 1]{Hughes}, we obtain for given $(\bx ,\bs, \bg)$ and covariance matrix $\cov(X,X',S|G=g)$ that the complement of each event in \eqref{eq:22Huz_c5}--\eqref{eq:21Huz_c4} happens with decreasingly doubly exponential probability for sufficiently large $n$ meaning that
\begin{align}
&\bbP\Big\{\!\big|\big\{\!m'\!:\!(\bx,\bx(m'),\bs,\bg)\!\in\!\T(X,X',S,G)\!\big\}\!\big| \!\le\! \exp\big\{n\big[|R-I(X';XSG)|^++\delta(\eps)\big]\big\}\Big\}\nonumber\\ &\hspace{26em} < \exp(-\exp(n\sigma(\eps))),\label{2exp_2}
\\
&\bbP\bigg\{\frac{1}{N}\big|\big\{\! m\!:\!(\bx(m),\bx(m'),\bs,\bg)\!\in\!\T\!(X,X'\!,S,G)\text{ for some }m'\!\ne\! m\!\big\}\big|\le \!2\exp\{\!-n\delta(\eps)/2\!\}\bigg\}\nonumber\\ & < \exp(-\exp(n\sigma(\eps))),\text{ if }I(X;X'SG)\!\ge\! |R\!-\! I(X';SG)|^+\!\!+\! \delta(\eps).\label{2exp_3}
\end{align}

Then, in order to complete the proof, since for any fixed $\nu$ the cardinality of finite set $\scU^n$ is only increasingly exponentially in $n$, and the set $\calV^{3\times 3}$ is finite along with the doubly decreasing exponential probabilities in \eqref{2exp_2}--\eqref{2exp_3}, we derive that with probability approaching to $1$, all inequalities in \eqref{eq:22Huz_c5}--\eqref{eq:21Huz_c4} hold simultaneously for sufficiently large $n$. Since these inequalities hold for every element in the finite sets $\scU^n$ and $\calV^{3\times 3}$, then for any vector $\bs, \bx$ and any given covariance matrix $\cov(X,X',S|G=g)$ (with $\|\bx\|^2= n, \|\bs\|^2\leq n\Lambda$) which is not in its corresponding $\nu$-dense subset, there exists a point in the corresponding $\nu$-dense subset that is close enough to it (in its $\nu$ distance neighborhood). Now, by using the continuity properties of all sets, we may conclude that \eqref{eq:22Huz_c5}--\eqref{eq:21Huz_c4} hold also for any point which is not in the $\nu$-dense subset. 

\bibliographystyle{IEEEtran}
\bibliography{mybib}

\end{document}